\documentclass[aps,twocolumn,showpacs,superscriptaddress,groupedaddress,floatfix]{revtex4}  
\usepackage{graphicx} 
\usepackage{dcolumn} 
\usepackage{bm}        
\usepackage{amssymb,amsmath}  
\usepackage{pdfpages}
\usepackage{multirow}
\usepackage{booktabs}
\usepackage{rotating}
\usepackage{capt-of}
\usepackage{hyperref}
\usepackage{siunitx}
\usepackage{url}
\usepackage{hyperref} \hypersetup{colorlinks=true,citecolor=blue,linkcolor=blue,urlcolor=blue} 
\usepackage{makeidx, multirow, longtable, tocbibind, amssymb}

\begin{document}

\title{On the Emergence of Topologically Protected Boundary States in Topological/Normal Insulator Heterostructures}

\author{M. Costa}
\affiliation{Brazilian Nanotechnology National Laboratory (LNNano), CNPEM, 13083-970 Campinas, Brazil}
\affiliation{Federal University of ABC, Santo Andr\'e, SP, Brazil}
\author{A. T. Costa}
\affiliation{Brazilian Nanotechnology National Laboratory (LNNano), CNPEM, 13083-970 Campinas, Brazil}
\affiliation{Departamento de F\'isica, Universidade Federal Fluminense, Niter\'oi, Rio de Janeiro, Brazil}
\author{Walter A. Freitas}
\author{Tome M. Schmidt}
\affiliation{Instituto de F\'isica, Universidade Federal de Uberl\^andia, CP 593, 38400-902, Uberl\^andia, MG, Brazil}
\author{M. Buongiorno Nardelli}
\affiliation{ Department of Physics and Department of Chemistry, University of North Texas, Denton TX, USA}
\affiliation{Center for Materials Genomics, Duke University, Durham, NC 27708, USA}
\author{A. Fazzio}
\affiliation{Brazilian Nanotechnology National Laboratory (LNNano), CNPEM, 13083-970 Campinas, Brazil}
\affiliation{Federal University of ABC, Santo Andr\'e, SP, Brazil}
\date{\today}

\begin{abstract}
We have performed a systematic investigation of the formation of topologically protected boundary states (TPBS) in topological/normal insulators (TI/NI) heterostructures. Using a recently developed scheme to construct {\it ab-initio} tight-binding Hamiltonian matrices from density functional theory (DFT) calculations,  we studied systems of realistic size with high accuracy and control over the relevant parameters such as TI and NI band alignment, NI gap and spin-orbit coupling strength. Our findings point to the existence of an NI critical thickness for the emergence of TPBS and to the importance of the band alignment between the TI and NI for the appearance of the TPBS. We chose Bi$_{2}$Se$_{3}$ as a prototypical case where the topological/normal insulator behavior is modeled by regions with/without spin-orbit coupling. Finally, we validate our approach comparing our model with fully relativistic DFT calculations for TI/NI heterostructures of Bi$_{2}$Se$_{3}$/Sb$_{2}$Se$_{3}$.

\end{abstract}

\pacs{}
\maketitle

\section{Introduction}

\begin{figure}
\includegraphics[scale=0.45,angle=-90]{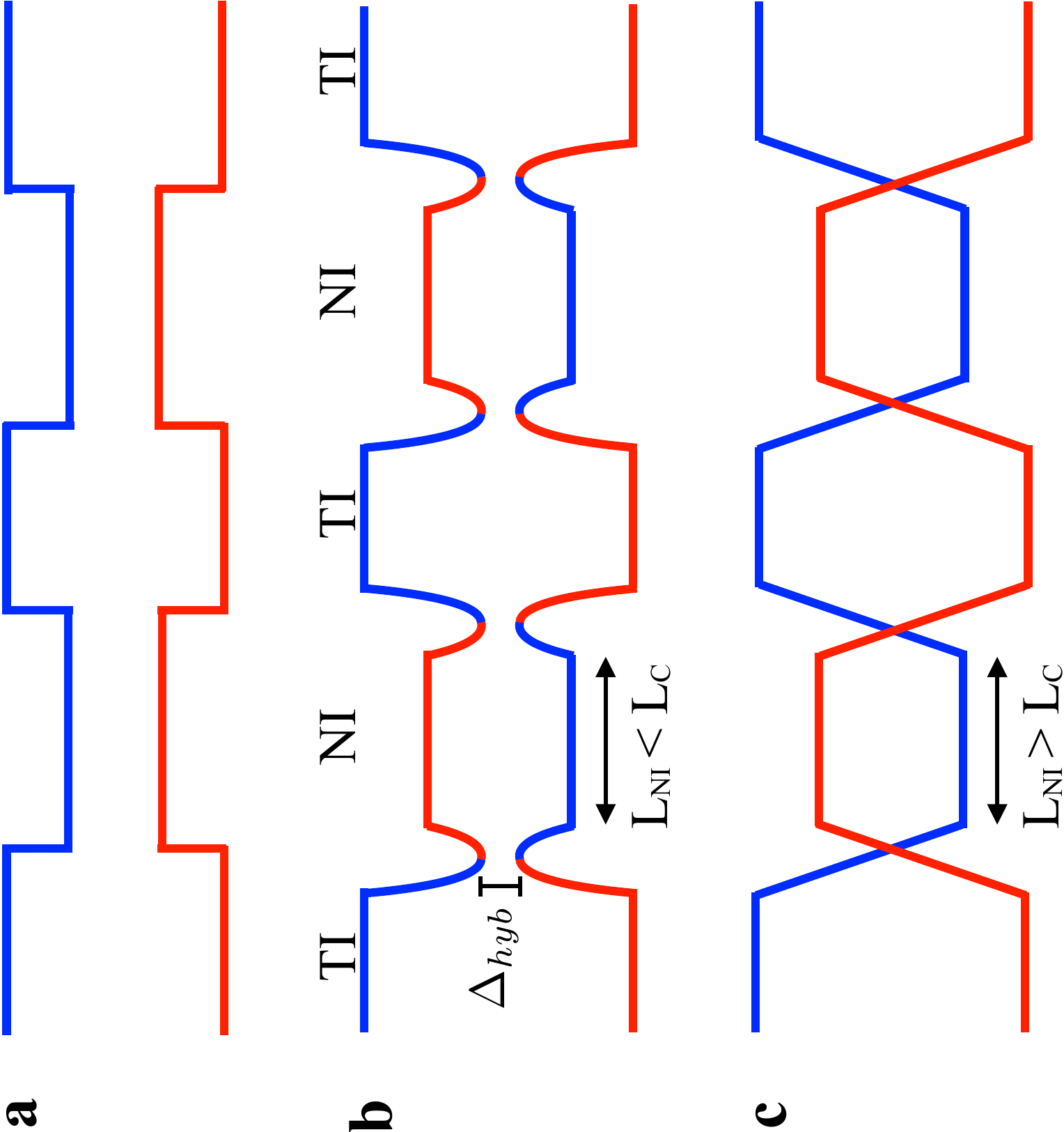}
\caption{\label{schematic} (color online) Schematic band structure for different heterostructures. (a) The regular semiconductor heterostructure. (b) Topological/normal insulator heterostructure (TI/NI) for NI thickness (L$_{NI}$) smaller than a critical value (L$_{C}$). The topologically protected boundary states (TPBS) are strongly hybridizing originating the $\Delta_{hyb}$ gap. (c) TI/NI heterostructure where L$_{NI}$ $>$ L$_{C}$ and the TPBS are decoupled.  }
\end{figure}

In 1970 Esaki and Tsui published a seminal paper where they proposed a solid-state artificial structure, the superlattice (SL)~\cite{SL}. This material is engineered as a periodic modulation of the composition of an alloy, with a period much larger than the host material lattice parameter (see Fig.~\ref{schematic} (a)). This new class of man-made semiconductor structures allowed a profusion of new technological applications to appear~\cite{SL3}. Nowadays SLs are not restricted to the realm of semiconductors but are ubiquitous throughout solid state physics. Different techniques are used in its synthesis, e.g., molecular beam epitaxy (MBE), metalorganic vapor phase epitaxy deposition (MOCVD). Furthermore, SLs are fabricated from a wide range of materials, such as magnetic and non-magnetic metals, insulators and superconductors~\cite{magnetic,superconductor}.

More recently, topological insulators (TI) ~\cite{TI1,TI2,TI3} were added to the previous list. TIs form a class of materials that present an insulating bulk with robust conducting states on its boundary. A question that arises naturally is how the TI properties are affected by interfacing with trivial materials. It is well known that on a TI surface the metallic topological states are Dirac-like with a spin texture that gives rise to a spin current protected from backscattering by time-reversal symmetry. Nevertheless, TIs interfaced with a regular or normal insulator (NI) are poorly understood. Some recent studies show unequivocally the appearance of Dirac cones at the TI/NI interface~\cite{interface1,interface2,interface3,interface4,Seixas2015,miwa,rwu,bi2se3-aln}. Existing calculations~\cite{proximity1,proximity2}, and recent experiments~\cite{proximity3} predict that states at the interface, located predominantly within the normal material, could acquire spin texture due to their proximity to the topological interface states. Now, let us consider a system composed of layers of TI and NI stacked in an alternating fashion, forming a one-dimensional superlattice. Such a system has been realized recently by stacking layers of pristine and In-doped Bi$_{2}$Se$_{3}$~\cite{hasan}, and ARPES measurements on such structure seem to suggest the emergence of a one-dimensional topological phase, drawn schematically in Fig.~\ref{schematic} (c).

Motivated by these experimental and theoretical results, we have performed a systematic study of the electronic and topological properties of heterostructures composed of topological and normal insulators. 

Our findings show a direct relationship between the normal insulator thickness (L$_{NI}$) and the appearance of topologically protected boundary states (TPBS). For a L$_{NI}$ smaller then a critical thickness (L$_{C}$) the heterostructure band structure will have the features of Fig.~\ref{schematic} (b), with an hybridization gap ($\Delta_{hyb}$) that is the result of TPBS interaction. For a TI/NI heterostructure, the TPBS is always present at the surface, independent of L$_{NI}$. Finally, our calculations point to an inverse relation between the NIs gap and its critical thickness. 

The paper is organized as follows: in Sec.~\ref{methodology} we give a brief description of the methodology used to construct the tight-binding Hamiltonians (TB) and how the spin-orbit coupling is included; we also confront these results to fully relativistic DFT calculations for validation. The computational details are described in sec.~\ref{comp-det}. In sec.~\ref{result-diss} the TB results are discussed and compared to DFT calculations. 

\begin{figure}
\includegraphics[scale=0.3,angle=-90]{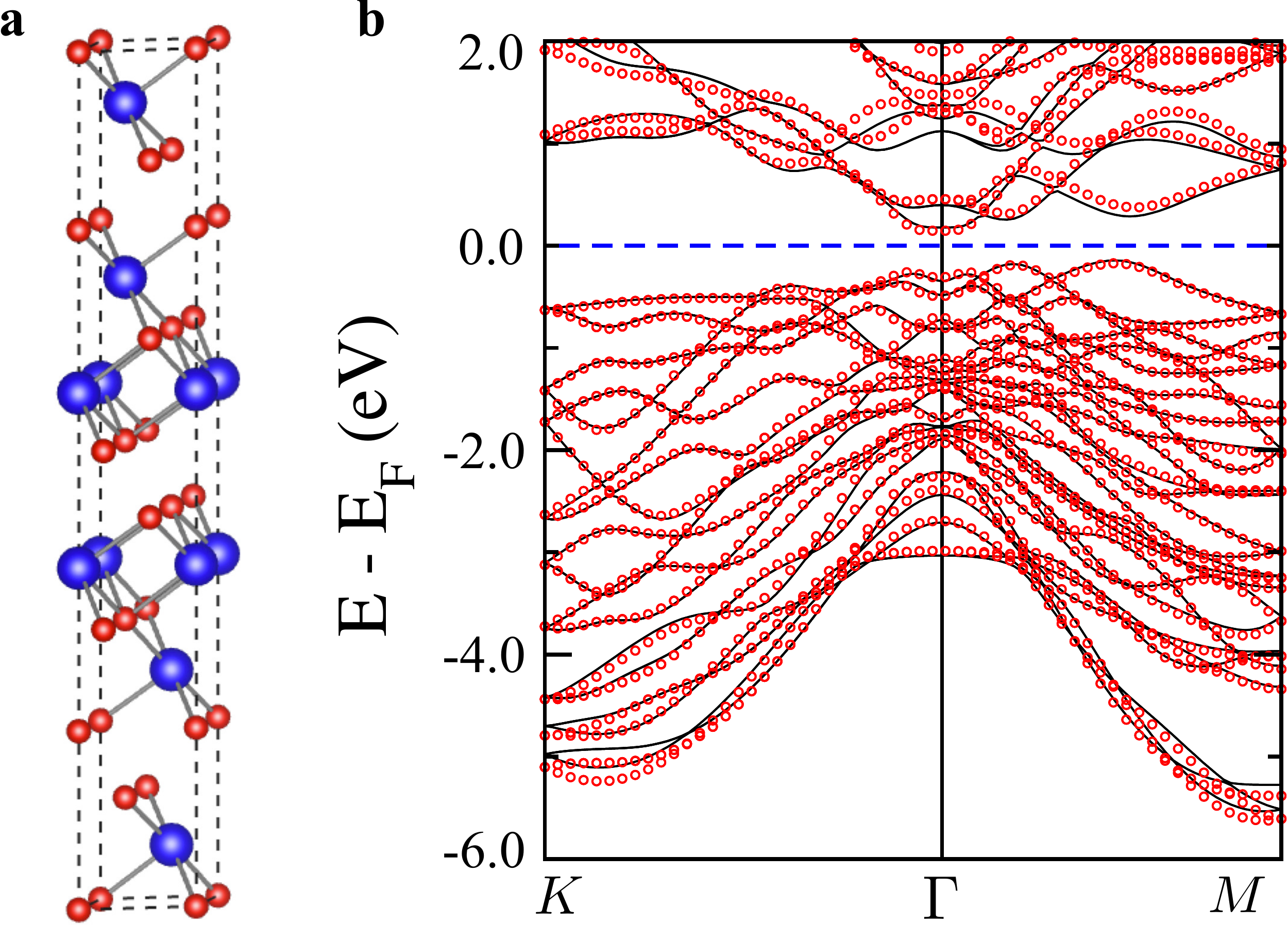}
\caption{\label{dftXtb-bulk} (color online) (a) Bi$_{2}$Se$_{3}$ bulk hexagonal crystal structure and its correspondent band structure (b)  ($K-\Gamma-M$)  for a fully relativistic DFT calculation (solid black line) compared to the TB+SOC (open red circles). The blue dashed line represents the Fermi level.}
\end{figure}

\section{Methodology}
\label{methodology}

\begin{figure}
\includegraphics[scale=0.43,angle=-90]{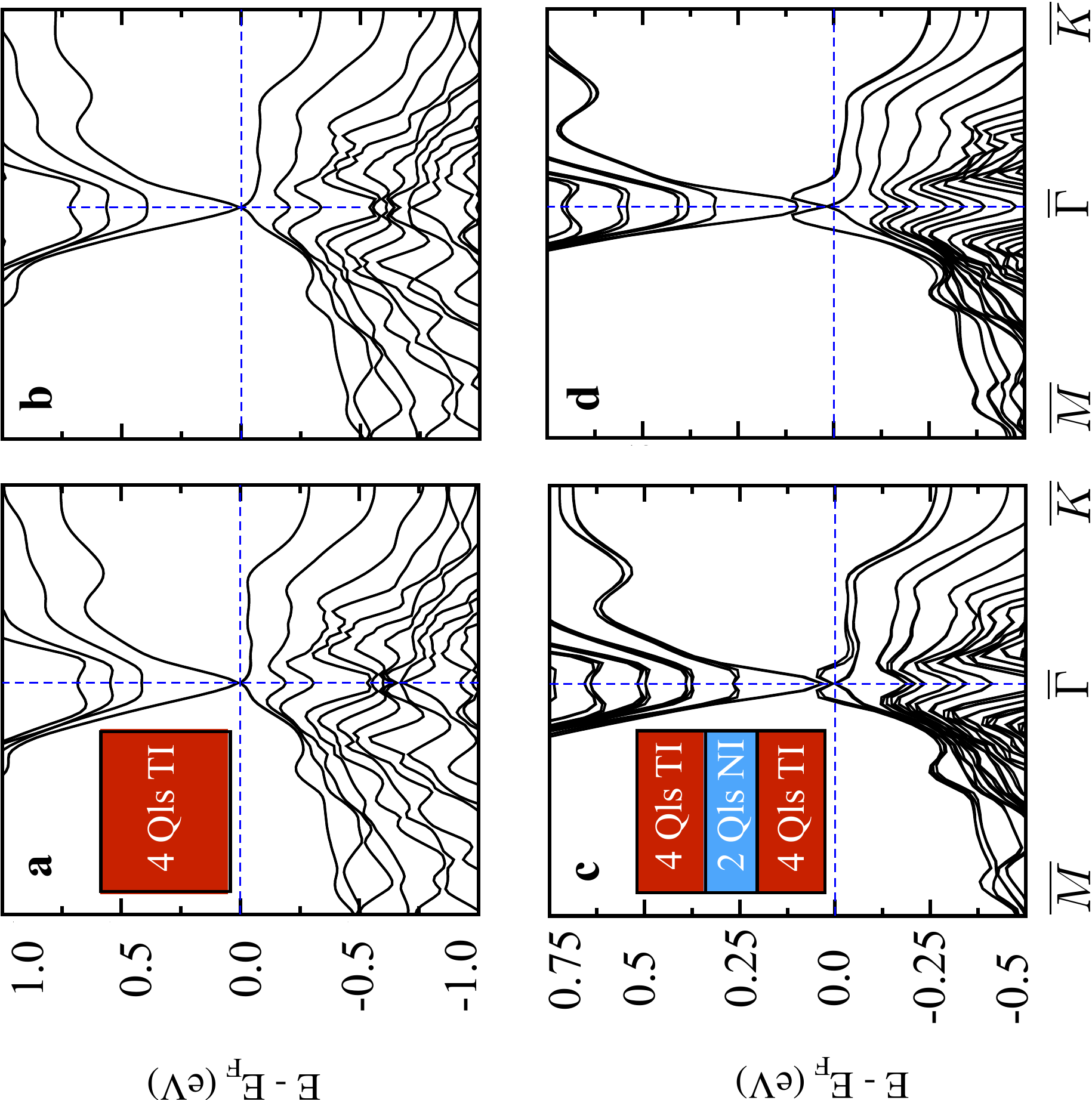}
\caption{\label{dftXtb} (color online) Comparison between the DFT-SOC and TB+SOC band structure. Bi$_{2}$Se$_{3}$ 4Qls surface band structure (a) DFT-SOC and (b) TB+SOC. 
Finite heterostructure with 2Qls of Bi$_{2}$Se$_{3}$ without SOC, normal insulator (NI), between two 4Qls of Bi$_{2}$Se$_{3}$ with SOC, topological insulator (TI), (c) DFT-SOC and (d) TB+SOC. The inset of panel (c) shows the schematic representation of the TI/NI/TI heterostructure. The red (blue) regions represent the TI (NI) layers.}
\end{figure}

This section briefly describes the methodology used for the tight-binding Hamiltonians construction and the inclusion of the spin-orbit coupling (SOC) effect. The central idea is to map the plane waves (PWs) Hilbert space, in general composed of several thousand of PWs, onto the compact sub-space spanned by the pseudo-atomic orbitals $|\phi_{\mu}\rangle$ (PAOs) used to construct the atomic pseudopotentials. The Hamiltonians are constructed from the Kohn-Sham (KS) Bloch states ($|\Psi^{KS}\rangle$) projection onto the PAO basis set localized at the atomic sites. Considering the PAO basis set incompleteness some KS states are not accurately described. This manifest itself as unphysical eigenvalues and hybridizations. The solution is to apply a filtering procedure to shift these states outside the energy window of interest. The filtering procedure is based on the so called projectability number ($p_{n}$), which is defined as $p_{n}=\langle\Psi_{n}|\hat{P}|\Psi_{n}\rangle$ and the operator $\hat{P}$ projects the KS states onto PAO basis set. The interpretation of $p_{n}$ is straightforward, KS states with $p_{n}$ $\approx$ 1 ($\approx$ 0) represents states which are well (poor) described by the given PAO basis set. In this work the $p_{n}$ threshold is set to 0.95. This methodology has been successful successfully applied to different materials and properties for a full description see references~\cite{PAO1,PAO2,PAO3,PAO4,PAO5}.

Spin-orbit coupling effect is essential to describe the topological quantum states in TI. In our calculations the SOC is introduced in two different ways: standard fully-relativistic self-consistent DFT calculations~\cite{DFT-SOC} and via an effective approximation in the TB Hamiltonian~\cite{soc,PAO5}. The effective SOC included in the TB Hamiltonian can be written as
\begin{equation}
H_{SOC} = \lambda {\bf L} . {\bf S}
\end{equation}
where $\lambda$ is the, orbital and element dependent, spin-orbit coupling strength, {\bf L} and {\bf S} are the orbital and spin angular momentum operators. Figure~\ref{dftXtb-bulk} shows a comparison between DFT with SOC (solid black line) and the TB+SOC (open red circles) in Bi$_{2}$Se$_{3}$ hexagonal bulk band structure. The Bi and Se $p$-orbitals SOC parameters used are $\lambda_{Bi}=1.475$ eV and $\lambda_{Se}=0.265$ eV. These values are in good agreement with the literature~\cite{bi2se3} and were obtained via a fitting procedure of the Bi and Se BCC DFT-SOC band structure. The agreement between the DFT and TB is excellent. Same degree of agreement is observed for a Bi$_{2}$Se$_{3}$ surface with 4 quintuple-layers (Qls), see Fig.~\ref{dftXtb} (a) and (b); and in TI/NI heterostructures: to this end we considered a heterostructure composed of 2 NI Qls between 4 TI Qls, see Fig.~\ref{dftXtb} (c) inset, using Bi$_{2}$Se$_{3}$ as the material of choice. In the TB Hamiltonian such heterostructure is constructed by setting Bi and Se SOC parameters to zero in the NI region, {\it i.e.} $\lambda_{Bi}=\lambda_{Se}=0$. The DFT calculation was performed with a Bi and Se non-relativistic pseudo potential in the NI region to mimic the SOC absence. Figure~\ref{dftXtb} (c) and (d) the DFT with SOC and TB+SOC heterostructure band-structure are showed, respectively. As in the bulk and surface cases the agreement is outstanding.

\begin{figure}
\includegraphics[scale=0.32,angle=-90]{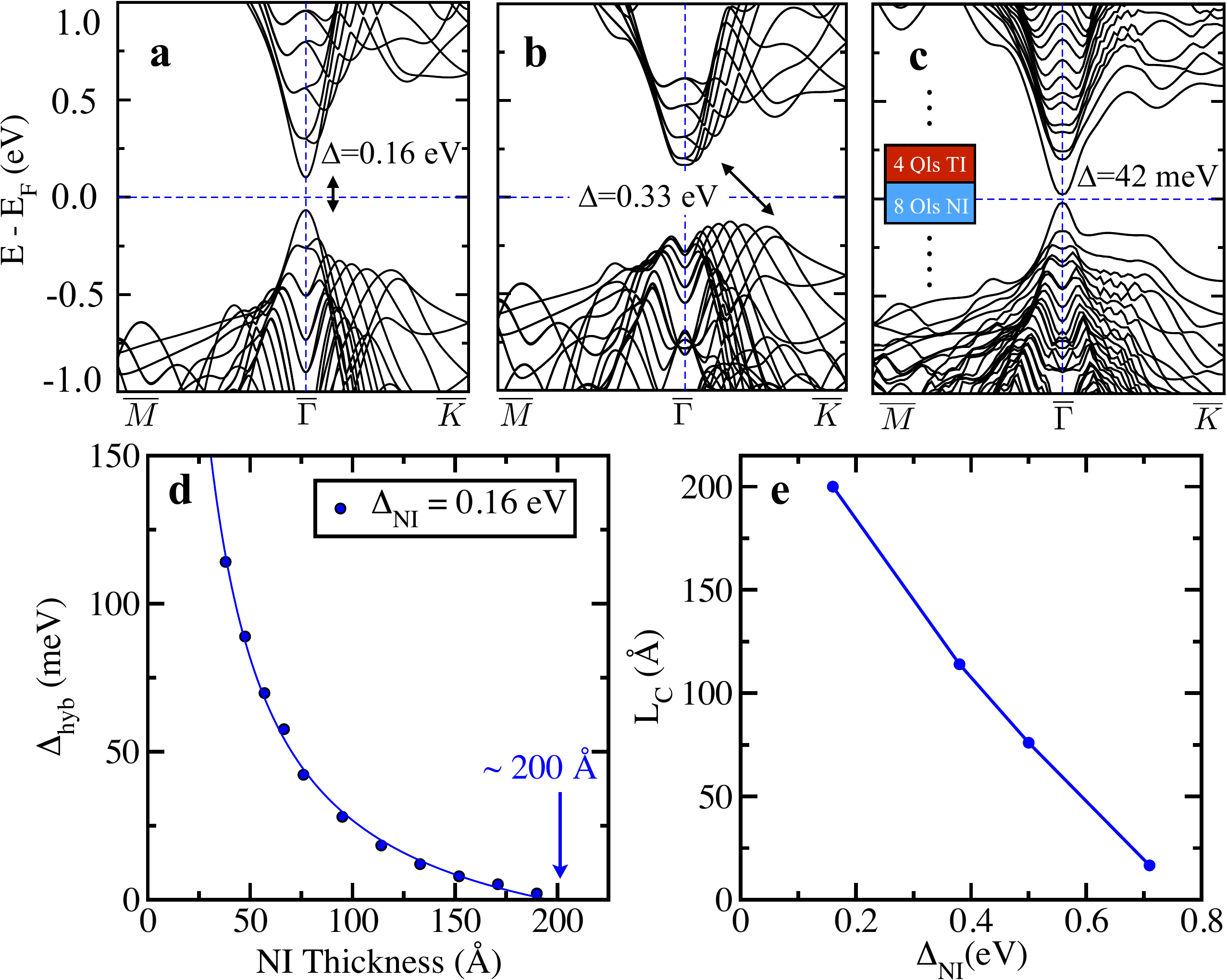}
\caption{\label{infinity} (color online) Bi$_{2}$Se$_{3}$ bulk band structure (a) without and (b) with SOC. (c) The 4/8 (TI/NI) heterostructure band structure, a 42 meV hybridization gap ($\Delta_{hyb}$) is obtained. (d) The $\Delta_{hyb}$ evolution as the NI thickness is increased. (e) The relation between the NI gap ($\Delta_{Nl}$) and the critical thickness (L$_{c}$) is shown.}
\end{figure}

\section{Computational Details}
\label{comp-det}

We performed Density Functional Theory (DFT)~\cite{DFT1,DFT2} calculations using the plane wave package Quantum Espresso (QE)~\cite{QE-2009,QE-2017}, and VASP code \cite{vasp1,vasp2}. The generalized gradient approximation (GGA)~\cite{pbe} was employed to treat the correlation among electrons with the Perdew-Burke-Ernzerhof (PBE) parametrization~\cite{pbe}. The Bi and Se Ionic potentials are described using PAW pseudo potentials~\cite{PAW} with $s$, $p$ and $d$ as valence. The Bi$_{2}$Se$_{3}$ is rhombohedral crystal with space group $D^{5}_{3d}$ ($R\overline{3}m$) and a five atoms unit cell. To facilitate the interface construction and results interpretation, a Bi$_{2}$Se$_{3}$ hexagonal cell composed of 15 atoms was used, see fig.~\ref{dftXtb-bulk} (a). This cell contains 3 Qls where the intra (inter) Ql chemical bonding is of ionic/covalent (van der Waals) character. The experimental lattice parameter is adopted with a=4.134 \AA~and c=28.63 \AA. The reciprocal space sampling uses the Monkhost-pack scheme with a k-mesh of 19$\times$19$\times$3 and 19$\times$19$\times$1 for bulk and surface calculations, respectively. We also performed a fully relativistic calculation for a heterostructure of Bi$_{2}$Se$_{3}$ and a real trivial system Sb$_{2}$Se$_{3}$. Since both systems present similar lattice parameters, we use the average lattice parameter for the hexagonal plane (a=4.107~{\AA}) and the vertical c=30.90~{\AA} for Sb$_{2}$Se$_{3}$. At the interface, the average of both structures was used c=29.765~{\AA}. The calculated topological invariant using these parameters gives $Z_2=1$ for  Bi$_{2}$Se$_{3}$ and $Z_2=0$ for Sb$_{2}$Se$_{3}$.
\section{Results and discussion}
\label{result-diss}

\begin{figure}
\includegraphics[scale=0.35,angle=-90]{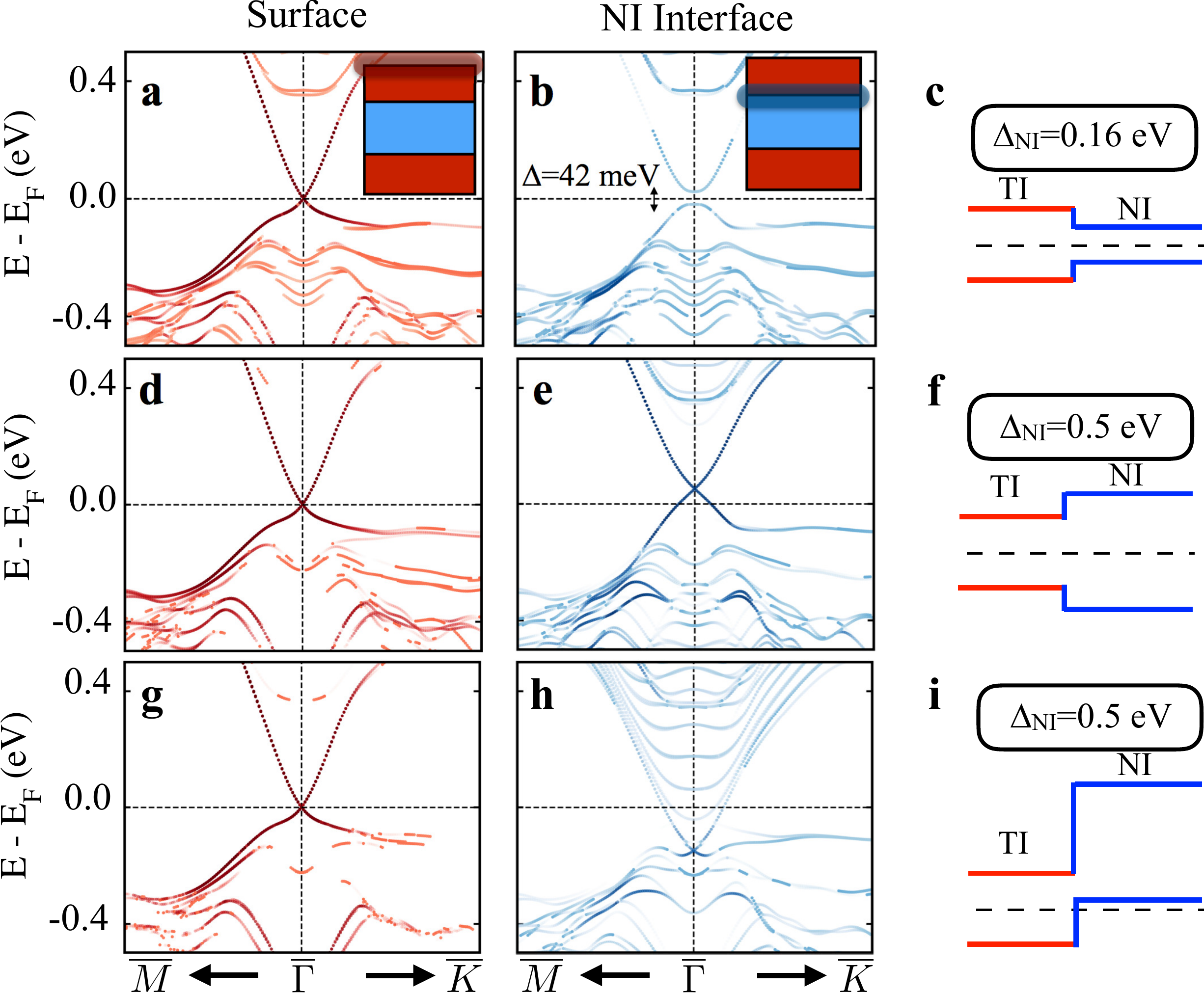}
\caption{\label{ti-ni-ti} (color online) Spatial projected band structure for a 4/8/4 (TI/NI/TI) finite heterostructure. Projection onto the TI outmost 2 Qls at the surface (a) $\Delta_{NI}$=0.16 eV (bands aligned), (b) $\Delta_{NI}$=0.50 eV (bands aligned) and (a) $\Delta_{NI}$=0.50 eV (bands not aligned) . In (b), (e) and (g) the projection is calculated at TI/NI interface for the respective heterostructure. The shaded area of panels (a) and (b) inset shows the projected regions. In (c), (f) and (i) a schematic representation of the band alignment for the respective heterostructure.}
\end{figure}

\begin{figure}
\includegraphics[scale=0.4,angle=-90]{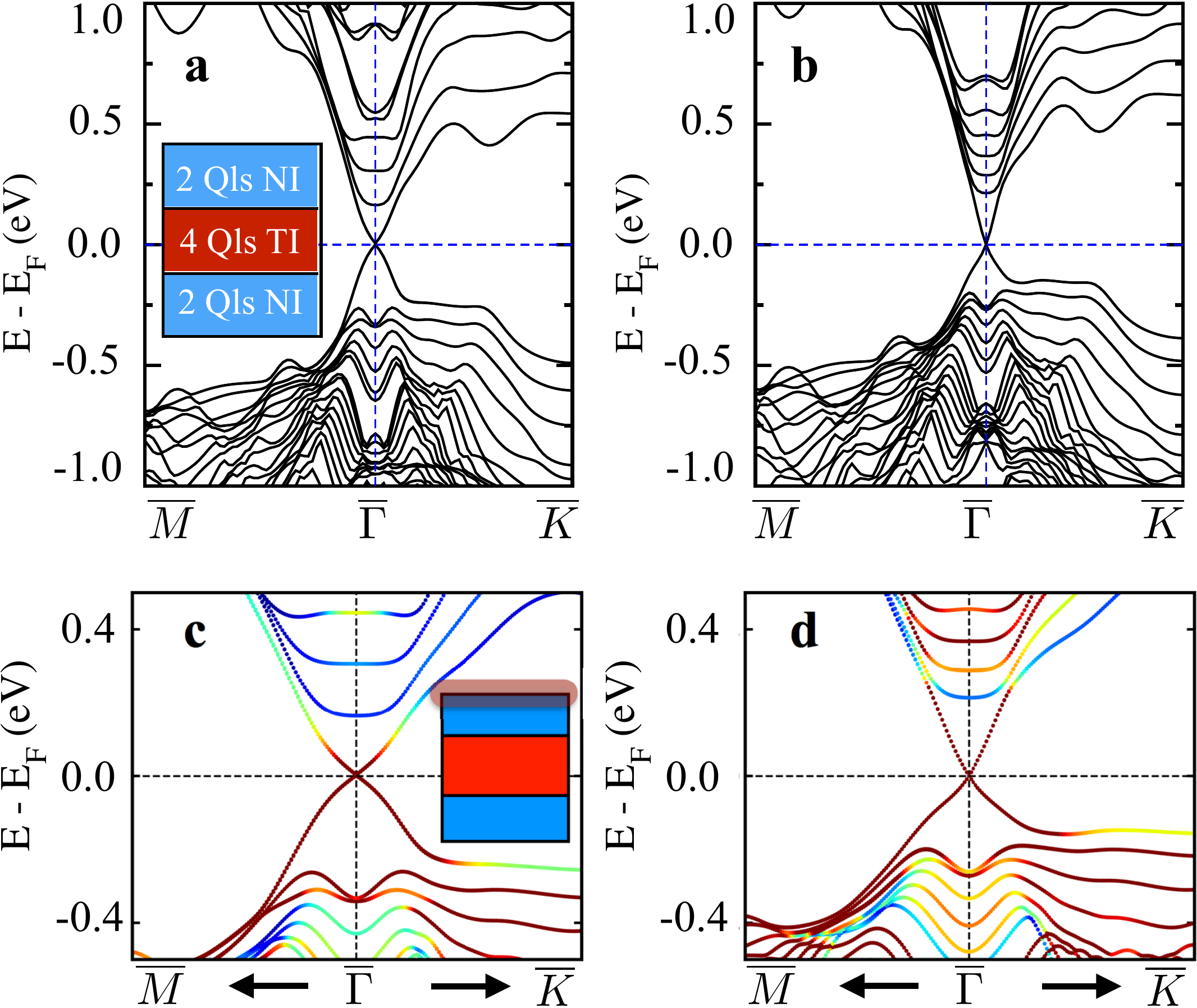}
\caption{\label{ni-ti-ni} (color online) Band structure for a 2/4/2 (NI/TI/NI) finite heterostructure. (a) $\lambda_{Bi}=0.0$ eV and (b) $\lambda_{Bi}=1.10$ eV. Eigenvalues spacial projection (c) $\lambda_{Bi}=0.0$ eV and (d) $\lambda_{Bi}=1.10$ eV. Dark red (dark blue) represents high (low) spectral weight at the projected regions. The inset of panel (c), shaded region, highlights the projected region.}
\end{figure}

\subsection{$\cdots$\{/TI/NI/\}$\cdots$ in TB}

In this section, we will establish the conditions for the appearance of the TPBS in infinite, perfectly periodic TI/NI heterostructures. In our calculations, the only difference between the NI and TI is the absence of SOC within the NI region. In Fig.~\ref{infinity} the Bi$_{2}$Se$_{3}$ bulk band structure is showed without SOC (a) and with SOC (b); the SOC increases the gap from 0.16 eV (direct gap) to 0.33 eV (indirect gap), which is in agreement with previous calculations~\cite{bi2se3}. In panel (c) we show the band structure for a 4/8 (TI/NI) infinite heterostructure where a hybridization gap ($\Delta_{hyb}$) of 42 meV appears at the $\overline{\Gamma}$ point. The $\Delta_{hyb}$ gap originates from the TPBS hybridization through the NI material. Its value is reduced by increasing the NI thickness and the TPBS reemerge above a critical thickness ($L_{c}$) around 200 \AA~($\approx$ 21 Qls), see panel (d). These superlattices are chemically and structural perfect; as a consequence, there is no Rashba splitting, which would appear due to inversion symmetry breaking, as reported in Ref.~\cite{NatHet}. In a real heterostructure the NI region is formed by different materials, thus it is important to understand the relation between the NI gap ($\Delta_{NI}$) and $L_{c}$. To increase $\Delta_{NI}$ a local potential is applied to Bi and Se $p$-states. There is a inverse relation between $\Delta_{NI}$ and L$_{c}$, see Fig.~\ref{infinity} (e). Increasing $\Delta_{NI}$ to 0.5 eV the L$_{c}$ is reduced to 75\AA~($\approx$ 8 Qls). This finding is important to guide the experimental search for suitable constituent materials of heterostructures with specific properties. 

\subsection{vacuum/TI/NI/TI/vacuum and vacuum/NI/TI/NI/vacuum in TB}

\begin{figure*}[ht]
\includegraphics[scale=0.65,angle=-90]{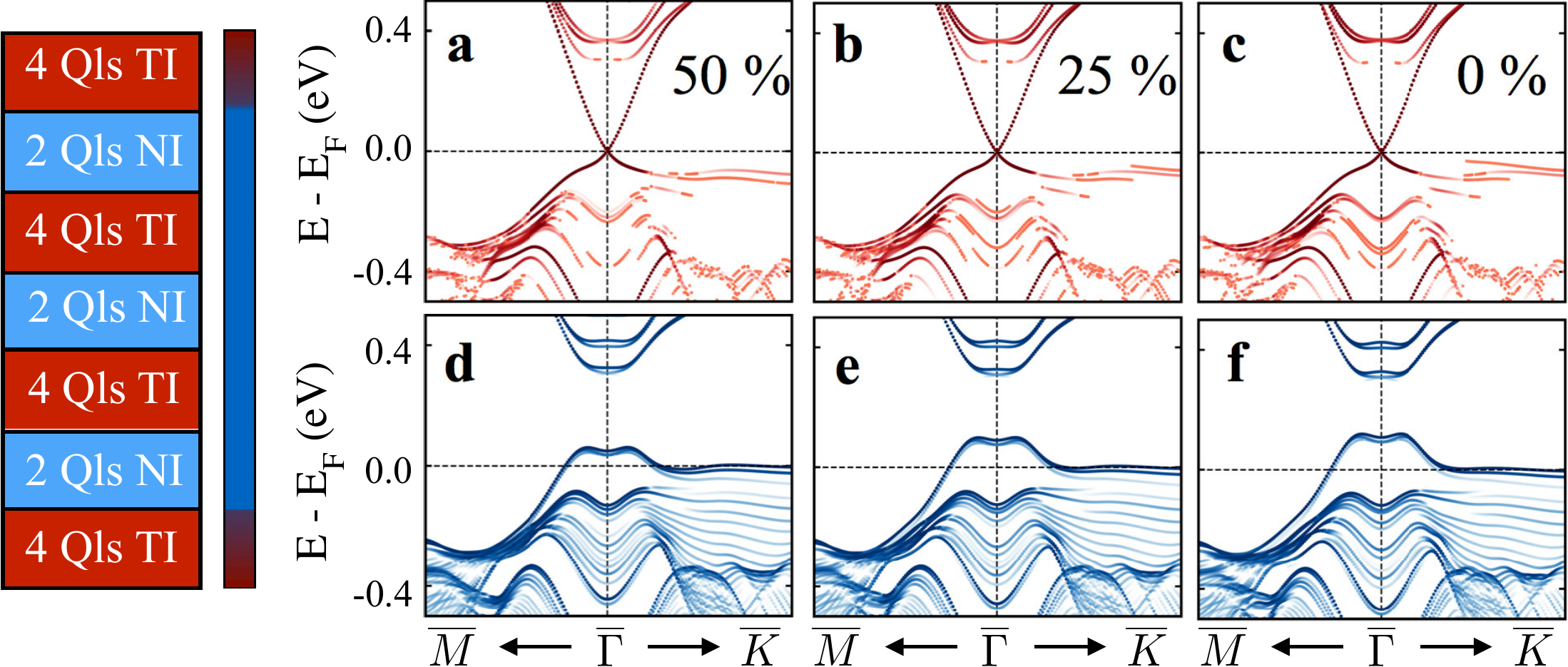}
\caption{\label{ti-ni-42} (color online) Band structure for a topological/normal insulator (TI/NI) finite heterostructure. The left panel shows the heterostructure schematic representation with 4 TI Qls (red) and 2 NI Qls (blue). Where in the NI region the Bi SOC strength is varied from 50\% of its full value to 0. The eigenvalues spacial projection onto the outmost 4 Qls TI, at the surface, is showed in (a) 50\%, (b) 25\% and (c) 0\% of the full Bi SOC. In (d), (e), and (f) the projection is onto the middle  region (TI and NI) of the slab. Dark red (dark blue) represents high (low) spectral weight at the projected region, see the colorbar.}
\end{figure*}

In this section we investigate heterostructures with vacuum terminations. Two different constructions will be explored:  one NI between two TI regions (TI/NI/TI), and one TI between two NI (NI/TI/NI) regions as building blocks for realistic TI/NI superlattices. In Fig.~\ref{ti-ni-ti} we show the band structure for the 4/8/4 (TI/NI/TI) sandwich, projected on the interfaces. The plot shows states with high (above 60\%) spectral weight at the indicated regions, inset of panels (a) and (b). In Fig.~\ref{ti-ni-ti} (a) and (b) the NI region gap is 0.16 eV (no potential is applied, and $\lambda_{Bi}=\lambda_{Se}=0.0$) and the TI and NI bands are perfectly aligned, as indicated in panel (c). The TPBS is observed at the surface, and their dispersions are consistent with the 4Qls Bi$_{2}$Se$_{3}$ surface. However, the TPBS at the TI/NI interface is replaced by states with parabolic dispersions and a 42 meV hybridization gap. This is in perfect agreement to the situation in the 4/8 infinite superlattice, including the size of the hybridization gap. In panels (d) and (e) we investigate the effect of increasing the NI band gap on the TPBS. For a gap of 0.5 eV, no significant change is observed at the surface. At the NI interfaces, however, the TPBS is restored, with an increased dispersion and an upwards energy shift. This result strongly suggests that large NI band gaps effectively decouple the two TI/NI interfaces. 
Similarly to ref.~\cite{band-alignment}, we explore the relation between band alignment and the TPBS formation. As in the previous cases, the surface TPBS is not significantly affected by the band misalignment (panel (g)) while the ones at the NI/TI interface are dramatically shifted downward in energy (panel (h)).

The next heterostructure is composed of a TI between two NI (NI/TI/NI) with equal thicknesses. With such construction, the TPBS's dispersion acquires a parabolic shape, as seen in Fig.~\ref{ni-ti-ni} (a). This larger dispersion results in the reduction of Fermi velocity when compared with the 4Qls surface linear dispersion, see Fig.~\ref{dftXtb} (b). Another important feature is the localization, in energy, of the bulk states between $\overline{\Gamma}$ and $\overline{M}$. For the 4Qls surface, this state is located 90 meV below the Dirac point (DP). Once the TI is placed in between two NI layers those states shift down to 240 meV below the DP. This effect is due to the absence of SOC at the surface since $\lambda_{Bi}=\lambda_{Se}=$0.0. As an exercise to understand this behavior the Se SOC is restored ($\lambda_{Se}=0.265$ eV) and the Bi SOC is increased to 1.10 eV, in the NI region. The bulk states are 150 meV below DP and move toward the surface value as $\lambda_{Bi}\rightarrow$ 100\%. This energy shift contributes to reduce the influence of these bulk states in the Bi$_{2}$Se$_{3}$ transport properties~\cite{bulk-states1,bulk-states2}. This has been observed for (Bi$_{x}$Sb$_{1-x}$)$_{2}$Se$_{3}$ alloys~\cite{abdalla}, where the Sb reduces the overall SOC. Moreover, panels (c) and (d) shows the spatial localization of the TPBS, a significant penetration of these states in the NI region is observed. This can be exploited in order to establish contacts for transport measurements or attain functionalization, for example. 

\begin{figure}[ht]
\includegraphics[scale=0.55,angle=-90]{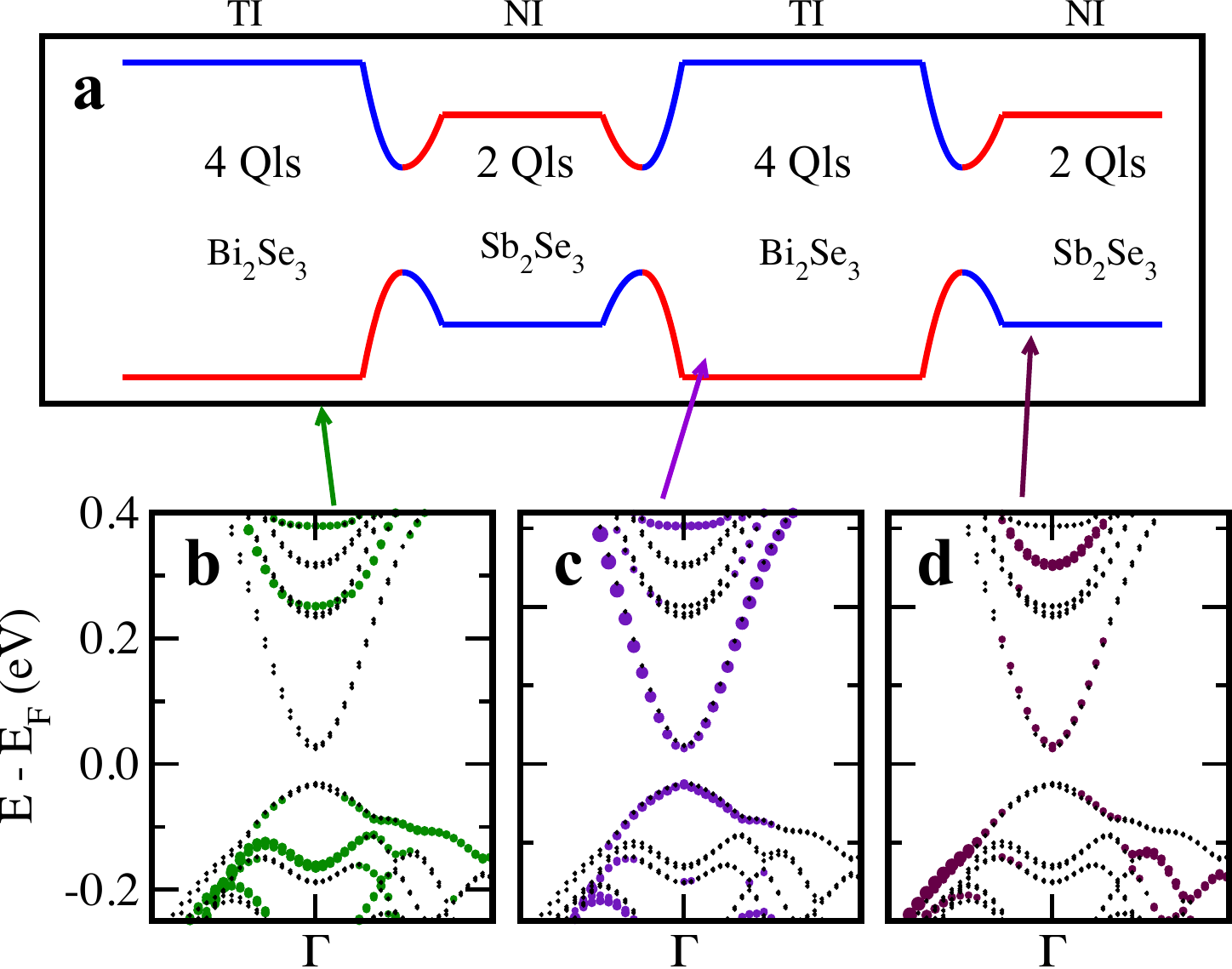}
\caption{\label{TI-NI-SL} (color online) (a) The TI/NI periodic superlattice schematic band structure. In (b), (c) and (d) is the spatial eigenvalue projection of a 1Ql from the TI center region, TI, and NI at the interface, respectively. }
\end{figure}

\subsection{$\cdots$~\{/TI/NI/\}$~\cdots$ in superlattice in TB}

We also investigate a superlattice composed by multilayers of topological and trivial insulators. In reference~\cite{hasan} this kind of multilayer has been presented as candidates to support a 1D chain of topological states. The modulation is performed by controlling the spin-orbit coupling, similarly to the control of topological phase by doping, as reported in ref.~\cite{hasan}. In our results, the intended 1D chain of topological states is not realized. At the TI surface the TPBS is always present. Nevertheless, at the TI/NI interface these states are absent due to a strong hybridization between subsequent interfaces. Figure~\ref{ti-ni-42}, left panel, shows the schematic representation for a 4/2 (TI/NI) heterostructure composed of 22 Qls ($\approx$ 200 \AA). In the TI region (red), full Bi and Se SOC is adopted. In the NI region (blue) the Se SOC is kept fixed at its full value and the Bi SOC is varied from 50 to 0\% of its full value. The Dirac-like states come from the TI in the surface (red in panels (a)-(c)), and no contributions come from the internal TI region (see panels (d)-(f)).  We can understand that by thinking about the 4/2 (TI/NI) structure as a unit cell of a ``new'', or ``meta'' material. This meta material is indeed a topological insulator with $Z_{2}=1$, since there is only one band inversion in the first Brillouin zone at the $\Gamma$ point. In this way, for such thin films of the TI/NI metamaterial we predict surface metallic topological states and a gapped bulk, exactly as shown in Fig.~\ref{ti-ni-42}. Naturally, by increasing the NI region greater than a critical thickness, as shown above, a perturbed gapless interface TPBS will be restored. However, in this case, one can not claim uncontroversially to have obtained a 1D chain of topological states since they are effectively decoupled from each other by the intervening NI layers.
\subsection{$\cdots$~\{/Bi$_{2}$Se$_{3}$/Sb$_{2}$Se$_{3}$/\}$~\cdots$ and vacuum/Bi$_{2}$Se$_{3}$/Sb$_{2}$Se$_{3}$/Bi$_{2}$Se$_{3}$/vacuum in DFT-SOC}
\begin{figure}[ht]
\includegraphics[scale=0.3,angle=-90]{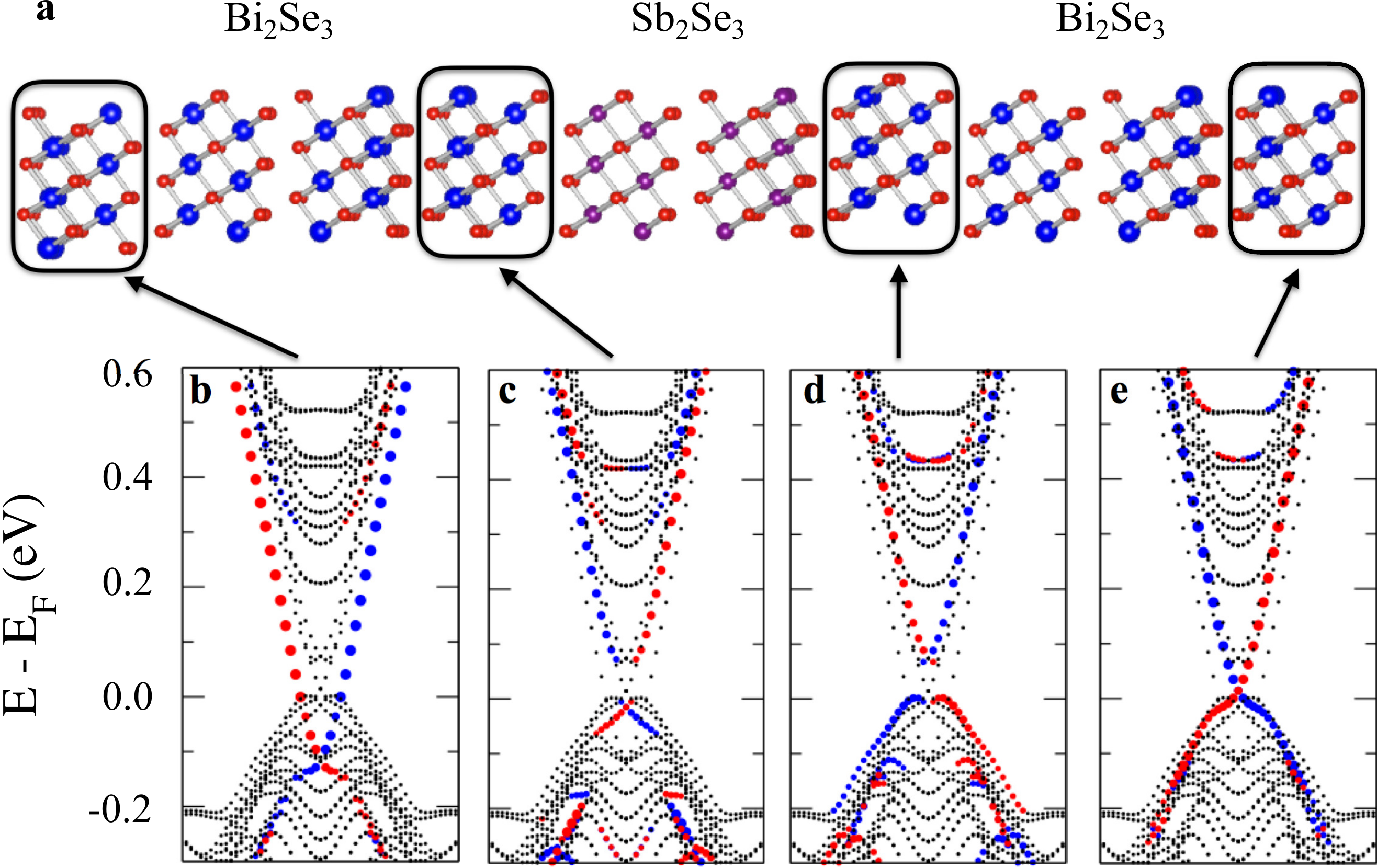}
\caption{\label{Slab} (color online) (a) 4/2/4 (TI/NI/TI) slab structure. In (b) - (e) are spin projections on 1st, 4th, 7th and 10th Ql, respectively.  Colors blue and red are projections of opposite in-plane spin texture. The Fermi level is at zero energy.}
\end{figure}

In order to complement our discussion above, fully relativistic calculations of the Bi$_{2}$Se$_{3}$ (TI) interfaced with Sb$_{2}$Se$_{3}$ (NI) were performed. Since both materials are lattice matched, the heterostructure preserves the bulk topological properties. For a periodic supercell composed of 4/2 (TI/NI), a gap is formed (see Fig.~\ref{TI-NI-SL}). There are topological surface bands at the interfaces which come mostly from the TI side, however, the global gap is preserved. By cutting our TI/NI supercell to produce top and bottom surfaces, a different picture is observed, confirming our TB results above. As shown in Fig.~\ref{Slab} The TPBS re-emerge at the surface and a gap is present at the Sb$_{2}$Se$_{3}$ interface. The whole system behaves as a unique TI material, gapped inside but metallic at the surfaces. As shown in Fig.~\ref{Slab} g), the TPBS are Rashba-like sates, due to the coupling between opposite interface TPBS, as observed for surface states in thin films \cite{Shan-2010}. The asymmetric distributions of the top and bottom surface states (Fig.~\ref{Slab} (f)) arise due to a symmetry breaking induced by the NI material. It breaks the Bi$_{2}$Se$_{3}$ symmetry since the unit cell demands 3 Qls, similar to stacking faults close to the surface \cite{Abdalla-2015}.

\section{Conclusions}
In conclusion, our calculations provide important insights to guide experimental realizations of TI/NI heterostructures with specific distributions of topologically protected boundary states (TPBS). Depending on the TI/NI heterostructure composition the TPBS can hybridize, through the intervening NI, leading to a bulk gap opening. We show that for NI layers thicker than a critical value (L$_{C}$) the hybridization is suppressed and the bulk TPBSs are restored. These TPBS are spatially localized at the TI/NI interface with a significant penetration in the NI region due to the proximity effect. Also, our calculations indicated an inverse relation between L$_{C}$ and the NI gap ($\Delta_{NI}$), suggesting that one should use a large gap trivial material to build the heterostructure if TPBSs in the internal TI/NII interfaces are desired. Moreover, the TI and NI band alignment will play an important role in the TPBS interface formation. With these results we expect to contribute to the understanding of the TI/NI interfacing problem which will be important in the device construction using topological insulators.
\section*{ACKNOWLEDGMENTS}
This work was supported by the Brazilian agencies FAPESP/TEM\'{ATICO} and CNPQ. We would like to acknowledge computing time provided by Laborat\'orio de Computa\c{c}\~ao  Cien\'ifica Avancada  (Universidade de S\~ao Paulo). We also like to thanks Prof. Yves Petroff for the fruitful discussion regarding ARPES measurements. 
\bibliographystyle{apsrev4-1} 
\normalbaselines
\bibliography{heterostructure}

\begin{thebibliography}{42}%
\makeatletter
\providecommand \@ifxundefined [1]{%
 \@ifx{#1\undefined}
}%
\providecommand \@ifnum [1]{%
 \ifnum #1\expandafter \@firstoftwo
 \else \expandafter \@secondoftwo
 \fi
}%
\providecommand \@ifx [1]{%
 \ifx #1\expandafter \@firstoftwo
 \else \expandafter \@secondoftwo
 \fi
}%
\providecommand \natexlab [1]{#1}%
\providecommand \enquote  [1]{``#1''}%
\providecommand \bibnamefont  [1]{#1}%
\providecommand \bibfnamefont [1]{#1}%
\providecommand \citenamefont [1]{#1}%
\providecommand \href@noop [0]{\@secondoftwo}%
\providecommand \href [0]{\begingroup \@sanitize@url \@href}%
\providecommand \@href[1]{\@@startlink{#1}\@@href}%
\providecommand \@@href[1]{\endgroup#1\@@endlink}%
\providecommand \@sanitize@url [0]{\catcode `\\12\catcode `\$12\catcode
  `\&12\catcode `\#12\catcode `\^12\catcode `\_12\catcode `\%12\relax}%
\providecommand \@@startlink[1]{}%
\providecommand \@@endlink[0]{}%
\providecommand \url  [0]{\begingroup\@sanitize@url \@url }%
\providecommand \@url [1]{\endgroup\@href {#1}{\urlprefix }}%
\providecommand \urlprefix  [0]{URL }%
\providecommand \Eprint [0]{\href }%
\providecommand \doibase [0]{http://dx.doi.org/}%
\providecommand \selectlanguage [0]{\@gobble}%
\providecommand \bibinfo  [0]{\@secondoftwo}%
\providecommand \bibfield  [0]{\@secondoftwo}%
\providecommand \translation [1]{[#1]}%
\providecommand \BibitemOpen [0]{}%
\providecommand \bibitemStop [0]{}%
\providecommand \bibitemNoStop [0]{.\EOS\space}%
\providecommand \EOS [0]{\spacefactor3000\relax}%
\providecommand \BibitemShut  [1]{\csname bibitem#1\endcsname}%
\let\auto@bib@innerbib\@empty
\bibitem [{\citenamefont {Esaki}\ and\ \citenamefont {Tsu}(1970)}]{SL}%
  \BibitemOpen
  \bibfield  {author} {\bibinfo {author} {\bibfnamefont {L.}~\bibnamefont
  {Esaki}}\ and\ \bibinfo {author} {\bibfnamefont {R.}~\bibnamefont {Tsu}},\
  }\href {http://ieeexplore.ieee.org/document/5391729/citations} {\bibfield
  {journal} {\bibinfo  {journal} {IBM J. Res. Dev.}\ }\textbf {\bibinfo
  {volume} {14}},\ \bibinfo {pages} {61} (\bibinfo {year} {1970})}\BibitemShut
  {NoStop}%
\bibitem [{SL3(1991)}]{SL3}%
  \BibitemOpen
  in\ \href@noop {} {\emph {\bibinfo {booktitle} {Quantum Semiconductor
  Structures}}},\ \bibinfo {editor} {edited by\ \bibinfo {editor}
  {\bibfnamefont {W.}~\bibnamefont {Claude}}\ and\ \bibinfo {editor}
  {\bibfnamefont {V.}~\bibnamefont {Borge}}}\ (\bibinfo  {publisher} {Academic
  Press},\ \bibinfo {address} {San Diego},\ \bibinfo {year} {1991})\BibitemShut
  {NoStop}%
\bibitem [{\citenamefont {Baibich}\ \emph {et~al.}(1988)\citenamefont
  {Baibich}, \citenamefont {Broto}, \citenamefont {Fert}, \citenamefont
  {Van~Dau}, \citenamefont {Petroff}, \citenamefont {Etienne}, \citenamefont
  {Creuzet}, \citenamefont {Friederich},\ and\ \citenamefont
  {Chazelas}}]{magnetic}%
  \BibitemOpen
  \bibfield  {author} {\bibinfo {author} {\bibfnamefont {M.~N.}\ \bibnamefont
  {Baibich}}, \bibinfo {author} {\bibfnamefont {J.~M.}\ \bibnamefont {Broto}},
  \bibinfo {author} {\bibfnamefont {A.}~\bibnamefont {Fert}}, \bibinfo {author}
  {\bibfnamefont {F.~N.}\ \bibnamefont {Van~Dau}}, \bibinfo {author}
  {\bibfnamefont {F.}~\bibnamefont {Petroff}}, \bibinfo {author} {\bibfnamefont
  {P.}~\bibnamefont {Etienne}}, \bibinfo {author} {\bibfnamefont
  {G.}~\bibnamefont {Creuzet}}, \bibinfo {author} {\bibfnamefont
  {A.}~\bibnamefont {Friederich}}, \ and\ \bibinfo {author} {\bibfnamefont
  {J.}~\bibnamefont {Chazelas}},\ }\href {\doibase 10.1103/PhysRevLett.61.2472}
  {\bibfield  {journal} {\bibinfo  {journal} {Phys. Rev. Lett.}\ }\textbf
  {\bibinfo {volume} {61}},\ \bibinfo {pages} {2472} (\bibinfo {year}
  {1988})}\BibitemShut {NoStop}%
\bibitem [{\citenamefont {Triscone}\ \emph {et~al.}(1990)\citenamefont
  {Triscone}, \citenamefont {Fischer}, \citenamefont {Brunner}, \citenamefont
  {Antognazza}, \citenamefont {Kent},\ and\ \citenamefont
  {Karkut}}]{superconductor}%
  \BibitemOpen
  \bibfield  {author} {\bibinfo {author} {\bibfnamefont {J.-M.}\ \bibnamefont
  {Triscone}}, \bibinfo {author} {\bibfnamefont {O.}~\bibnamefont {Fischer}},
  \bibinfo {author} {\bibfnamefont {O.}~\bibnamefont {Brunner}}, \bibinfo
  {author} {\bibfnamefont {L.}~\bibnamefont {Antognazza}}, \bibinfo {author}
  {\bibfnamefont {A.~D.}\ \bibnamefont {Kent}}, \ and\ \bibinfo {author}
  {\bibfnamefont {M.~G.}\ \bibnamefont {Karkut}},\ }\href {\doibase
  10.1103/PhysRevLett.64.804} {\bibfield  {journal} {\bibinfo  {journal} {Phys.
  Rev. Lett.}\ }\textbf {\bibinfo {volume} {64}},\ \bibinfo {pages} {804}
  (\bibinfo {year} {1990})}\BibitemShut {NoStop}%
\bibitem [{\citenamefont {Kane}\ and\ \citenamefont {Mele}(2005)}]{TI1}%
  \BibitemOpen
  \bibfield  {author} {\bibinfo {author} {\bibfnamefont {C.~L.}\ \bibnamefont
  {Kane}}\ and\ \bibinfo {author} {\bibfnamefont {E.~J.}\ \bibnamefont
  {Mele}},\ }\href {\doibase 10.1103/PhysRevLett.95.146802} {\bibfield
  {journal} {\bibinfo  {journal} {Phys. Rev. Lett.}\ }\textbf {\bibinfo
  {volume} {95}},\ \bibinfo {pages} {146802} (\bibinfo {year}
  {2005})}\BibitemShut {NoStop}%
\bibitem [{\citenamefont {Bernevig}\ \emph {et~al.}(2006)\citenamefont
  {Bernevig}, \citenamefont {Hughes},\ and\ \citenamefont {Zhang}}]{TI2}%
  \BibitemOpen
  \bibfield  {author} {\bibinfo {author} {\bibfnamefont {B.~A.}\ \bibnamefont
  {Bernevig}}, \bibinfo {author} {\bibfnamefont {T.~L.}\ \bibnamefont
  {Hughes}}, \ and\ \bibinfo {author} {\bibfnamefont {S.-C.}\ \bibnamefont
  {Zhang}},\ }\href {\doibase 10.1126/science.1133734} {\bibfield  {journal}
  {\bibinfo  {journal} {Science}\ }\textbf {\bibinfo {volume} {314}},\ \bibinfo
  {pages} {1757} (\bibinfo {year} {2006})}\BibitemShut {NoStop}%
\bibitem [{\citenamefont {Fu}\ \emph {et~al.}(2007)\citenamefont {Fu},
  \citenamefont {Kane},\ and\ \citenamefont {Mele}}]{TI3}%
  \BibitemOpen
  \bibfield  {author} {\bibinfo {author} {\bibfnamefont {L.}~\bibnamefont
  {Fu}}, \bibinfo {author} {\bibfnamefont {C.~L.}\ \bibnamefont {Kane}}, \ and\
  \bibinfo {author} {\bibfnamefont {E.~J.}\ \bibnamefont {Mele}},\ }\href
  {\doibase 10.1103/PhysRevLett.98.106803} {\bibfield  {journal} {\bibinfo
  {journal} {Phys. Rev. Lett.}\ }\textbf {\bibinfo {volume} {98}},\ \bibinfo
  {pages} {106803} (\bibinfo {year} {2007})}\BibitemShut {NoStop}%
\bibitem [{\citenamefont {Hirahara}\ \emph {et~al.}(2011)\citenamefont
  {Hirahara}, \citenamefont {Bihlmayer}, \citenamefont {Sakamoto},
  \citenamefont {Yamada}, \citenamefont {Miyazaki}, \citenamefont {Kimura},
  \citenamefont {Bl\"ugel},\ and\ \citenamefont {Hasegawa}}]{interface1}%
  \BibitemOpen
  \bibfield  {author} {\bibinfo {author} {\bibfnamefont {T.}~\bibnamefont
  {Hirahara}}, \bibinfo {author} {\bibfnamefont {G.}~\bibnamefont {Bihlmayer}},
  \bibinfo {author} {\bibfnamefont {Y.}~\bibnamefont {Sakamoto}}, \bibinfo
  {author} {\bibfnamefont {M.}~\bibnamefont {Yamada}}, \bibinfo {author}
  {\bibfnamefont {H.}~\bibnamefont {Miyazaki}}, \bibinfo {author}
  {\bibfnamefont {S.-i.}\ \bibnamefont {Kimura}}, \bibinfo {author}
  {\bibfnamefont {S.}~\bibnamefont {Bl\"ugel}}, \ and\ \bibinfo {author}
  {\bibfnamefont {S.}~\bibnamefont {Hasegawa}},\ }\href {\doibase
  10.1103/PhysRevLett.107.166801} {\bibfield  {journal} {\bibinfo  {journal}
  {Phys. Rev. Lett.}\ }\textbf {\bibinfo {volume} {107}},\ \bibinfo {pages}
  {166801} (\bibinfo {year} {2011})}\BibitemShut {NoStop}%
\bibitem [{\citenamefont {Berntsen}\ \emph {et~al.}(2013)\citenamefont
  {Berntsen}, \citenamefont {G\"otberg}, \citenamefont {Wojek},\ and\
  \citenamefont {Tjernberg}}]{interface2}%
  \BibitemOpen
  \bibfield  {author} {\bibinfo {author} {\bibfnamefont {M.~H.}\ \bibnamefont
  {Berntsen}}, \bibinfo {author} {\bibfnamefont {O.}~\bibnamefont {G\"otberg}},
  \bibinfo {author} {\bibfnamefont {B.~M.}\ \bibnamefont {Wojek}}, \ and\
  \bibinfo {author} {\bibfnamefont {O.}~\bibnamefont {Tjernberg}},\ }\href
  {\doibase 10.1103/PhysRevB.88.195132} {\bibfield  {journal} {\bibinfo
  {journal} {Phys. Rev. B}\ }\textbf {\bibinfo {volume} {88}},\ \bibinfo
  {pages} {195132} (\bibinfo {year} {2013})}\BibitemShut {NoStop}%
\bibitem [{\citenamefont {Lee}\ and\ \citenamefont
  {Yazyev}(2017)}]{interface3}%
  \BibitemOpen
  \bibfield  {author} {\bibinfo {author} {\bibfnamefont {H.}~\bibnamefont
  {Lee}}\ and\ \bibinfo {author} {\bibfnamefont {O.~V.}\ \bibnamefont
  {Yazyev}},\ }\href {\doibase 10.1103/PhysRevB.95.085304} {\bibfield
  {journal} {\bibinfo  {journal} {Phys. Rev. B}\ }\textbf {\bibinfo {volume}
  {95}},\ \bibinfo {pages} {085304} (\bibinfo {year} {2017})}\BibitemShut
  {NoStop}%
\bibitem [{\citenamefont {Takahashi}\ and\ \citenamefont
  {Murakami}(2011)}]{interface4}%
  \BibitemOpen
  \bibfield  {author} {\bibinfo {author} {\bibfnamefont {R.}~\bibnamefont
  {Takahashi}}\ and\ \bibinfo {author} {\bibfnamefont {S.}~\bibnamefont
  {Murakami}},\ }\href {\doibase 10.1103/PhysRevLett.107.166805} {\bibfield
  {journal} {\bibinfo  {journal} {Phys. Rev. Lett.}\ }\textbf {\bibinfo
  {volume} {107}},\ \bibinfo {pages} {166805} (\bibinfo {year}
  {2011})}\BibitemShut {NoStop}%
\bibitem [{\citenamefont {Seixas}\ \emph {et~al.}(2015)\citenamefont {Seixas},
  \citenamefont {West}, \citenamefont {Fazzio},\ and\ \citenamefont
  {Zhang}}]{Seixas2015}%
  \BibitemOpen
  \bibfield  {author} {\bibinfo {author} {\bibfnamefont {L.}~\bibnamefont
  {Seixas}}, \bibinfo {author} {\bibfnamefont {D.}~\bibnamefont {West}},
  \bibinfo {author} {\bibfnamefont {A.}~\bibnamefont {Fazzio}}, \ and\ \bibinfo
  {author} {\bibfnamefont {S.~B.}\ \bibnamefont {Zhang}},\ }\href
  {http://dx.doi.org/10.1038/ncomms8630} {\bibfield  {journal} {\bibinfo
  {journal} {Nature Communications}\ }\textbf {\bibinfo {volume} {6}},\
  \bibinfo {pages} {7630 EP } (\bibinfo {year} {2015})},\ \bibinfo {note}
  {article}\BibitemShut {NoStop}%
\bibitem [{\citenamefont {de~Oliveira}\ and\ \citenamefont
  {Miwa}(2016)}]{miwa}%
  \BibitemOpen
  \bibfield  {author} {\bibinfo {author} {\bibfnamefont {I.~S.~S.}\
  \bibnamefont {de~Oliveira}}\ and\ \bibinfo {author} {\bibfnamefont {R.~H.}\
  \bibnamefont {Miwa}},\ }\href
  {http://stacks.iop.org/0957-4484/27/i=3/a=035704} {\bibfield  {journal}
  {\bibinfo  {journal} {Nanotechnology}\ }\textbf {\bibinfo {volume} {27}},\
  \bibinfo {pages} {035704} (\bibinfo {year} {2016})}\BibitemShut {NoStop}%
\bibitem [{\citenamefont {Kim}\ \emph {et~al.}(2017)\citenamefont {Kim},
  \citenamefont {Kim}, \citenamefont {Wang}, \citenamefont {Sinova},\ and\
  \citenamefont {Wu}}]{rwu}%
  \BibitemOpen
  \bibfield  {author} {\bibinfo {author} {\bibfnamefont {J.}~\bibnamefont
  {Kim}}, \bibinfo {author} {\bibfnamefont {K.-W.}\ \bibnamefont {Kim}},
  \bibinfo {author} {\bibfnamefont {H.}~\bibnamefont {Wang}}, \bibinfo {author}
  {\bibfnamefont {J.}~\bibnamefont {Sinova}}, \ and\ \bibinfo {author}
  {\bibfnamefont {R.}~\bibnamefont {Wu}},\ }\href {\doibase
  10.1103/PhysRevLett.119.027201} {\bibfield  {journal} {\bibinfo  {journal}
  {Phys. Rev. Lett.}\ }\textbf {\bibinfo {volume} {119}},\ \bibinfo {pages}
  {027201} (\bibinfo {year} {2017})}\BibitemShut {NoStop}%
\bibitem [{\citenamefont {Freitas}\ \emph {et~al.}(2016)\citenamefont
  {Freitas}, \citenamefont {Fazzio},\ and\ \citenamefont
  {Schmidt}}]{bi2se3-aln}%
  \BibitemOpen
  \bibfield  {author} {\bibinfo {author} {\bibfnamefont {W.~A.}\ \bibnamefont
  {Freitas}}, \bibinfo {author} {\bibfnamefont {A.}~\bibnamefont {Fazzio}}, \
  and\ \bibinfo {author} {\bibfnamefont {T.~M.}\ \bibnamefont {Schmidt}},\
  }\href {\doibase 10.1063/1.4963350} {\bibfield  {journal} {\bibinfo
  {journal} {Applied Physics Letters}\ }\textbf {\bibinfo {volume} {109}},\
  \bibinfo {pages} {131601} (\bibinfo {year} {2016})},\ \Eprint
  {http://arxiv.org/abs/https://doi.org/10.1063/1.4963350}
  {https://doi.org/10.1063/1.4963350} \BibitemShut {NoStop}%
\bibitem [{\citenamefont {Hutasoit}\ and\ \citenamefont
  {Stanescu}(2011)}]{proximity1}%
  \BibitemOpen
  \bibfield  {author} {\bibinfo {author} {\bibfnamefont {J.~A.}\ \bibnamefont
  {Hutasoit}}\ and\ \bibinfo {author} {\bibfnamefont {T.~D.}\ \bibnamefont
  {Stanescu}},\ }\href {\doibase 10.1103/PhysRevB.84.085103} {\bibfield
  {journal} {\bibinfo  {journal} {Phys. Rev. B}\ }\textbf {\bibinfo {volume}
  {84}},\ \bibinfo {pages} {085103} (\bibinfo {year} {2011})}\BibitemShut
  {NoStop}%
\bibitem [{\citenamefont {Zhang}\ \emph {et~al.}(2014)\citenamefont {Zhang},
  \citenamefont {Triola},\ and\ \citenamefont {Rossi}}]{proximity2}%
  \BibitemOpen
  \bibfield  {author} {\bibinfo {author} {\bibfnamefont {J.}~\bibnamefont
  {Zhang}}, \bibinfo {author} {\bibfnamefont {C.}~\bibnamefont {Triola}}, \
  and\ \bibinfo {author} {\bibfnamefont {E.}~\bibnamefont {Rossi}},\ }\href
  {\doibase 10.1103/PhysRevLett.112.096802} {\bibfield  {journal} {\bibinfo
  {journal} {Phys. Rev. Lett.}\ }\textbf {\bibinfo {volume} {112}},\ \bibinfo
  {pages} {096802} (\bibinfo {year} {2014})}\BibitemShut {NoStop}%
\bibitem [{\citenamefont {Shoman}\ \emph {et~al.}(2015)\citenamefont {Shoman},
  \citenamefont {Takayama}, \citenamefont {Sato}, \citenamefont {Souma},
  \citenamefont {Takahashi}, \citenamefont {Oguchi}, \citenamefont {Segawa},\
  and\ \citenamefont {Ando}}]{proximity3}%
  \BibitemOpen
  \bibfield  {author} {\bibinfo {author} {\bibfnamefont {T.}~\bibnamefont
  {Shoman}}, \bibinfo {author} {\bibfnamefont {A.}~\bibnamefont {Takayama}},
  \bibinfo {author} {\bibfnamefont {T.}~\bibnamefont {Sato}}, \bibinfo {author}
  {\bibfnamefont {S.}~\bibnamefont {Souma}}, \bibinfo {author} {\bibfnamefont
  {T.}~\bibnamefont {Takahashi}}, \bibinfo {author} {\bibfnamefont
  {T.}~\bibnamefont {Oguchi}}, \bibinfo {author} {\bibfnamefont
  {K.}~\bibnamefont {Segawa}}, \ and\ \bibinfo {author} {\bibfnamefont
  {Y.}~\bibnamefont {Ando}},\ }\href {http://dx.doi.org/10.1038/ncomms7547}
  {\bibfield  {journal} {\bibinfo  {journal} {Nature Communications}\ }\textbf
  {\bibinfo {volume} {6}},\ \bibinfo {pages} {6547 EP } (\bibinfo {year}
  {2015})},\ \bibinfo {note} {article}\BibitemShut {NoStop}%
\bibitem [{\citenamefont {Belopolski}\ \emph {et~al.}(2017)\citenamefont
  {Belopolski}, \citenamefont {Xu}, \citenamefont {Koirala}, \citenamefont
  {Liu}, \citenamefont {Bian}, \citenamefont {Strocov}, \citenamefont {Chang},
  \citenamefont {Neupane}, \citenamefont {Alidoust}, \citenamefont {Sanchez},
  \citenamefont {Zheng}, \citenamefont {Brahlek}, \citenamefont {Rogalev},
  \citenamefont {Kim}, \citenamefont {Plumb}, \citenamefont {Chen},
  \citenamefont {Bertran}, \citenamefont {Le~F{\`e}vre}, \citenamefont
  {Taleb-Ibrahimi}, \citenamefont {Asensio}, \citenamefont {Shi}, \citenamefont
  {Lin}, \citenamefont {Hoesch}, \citenamefont {Oh},\ and\ \citenamefont
  {Hasan}}]{hasan}%
  \BibitemOpen
  \bibfield  {author} {\bibinfo {author} {\bibfnamefont {I.}~\bibnamefont
  {Belopolski}}, \bibinfo {author} {\bibfnamefont {S.-Y.}\ \bibnamefont {Xu}},
  \bibinfo {author} {\bibfnamefont {N.}~\bibnamefont {Koirala}}, \bibinfo
  {author} {\bibfnamefont {C.}~\bibnamefont {Liu}}, \bibinfo {author}
  {\bibfnamefont {G.}~\bibnamefont {Bian}}, \bibinfo {author} {\bibfnamefont
  {V.~N.}\ \bibnamefont {Strocov}}, \bibinfo {author} {\bibfnamefont
  {G.}~\bibnamefont {Chang}}, \bibinfo {author} {\bibfnamefont
  {M.}~\bibnamefont {Neupane}}, \bibinfo {author} {\bibfnamefont
  {N.}~\bibnamefont {Alidoust}}, \bibinfo {author} {\bibfnamefont
  {D.}~\bibnamefont {Sanchez}}, \bibinfo {author} {\bibfnamefont
  {H.}~\bibnamefont {Zheng}}, \bibinfo {author} {\bibfnamefont
  {M.}~\bibnamefont {Brahlek}}, \bibinfo {author} {\bibfnamefont
  {V.}~\bibnamefont {Rogalev}}, \bibinfo {author} {\bibfnamefont
  {T.}~\bibnamefont {Kim}}, \bibinfo {author} {\bibfnamefont {N.~C.}\
  \bibnamefont {Plumb}}, \bibinfo {author} {\bibfnamefont {C.}~\bibnamefont
  {Chen}}, \bibinfo {author} {\bibfnamefont {F.}~\bibnamefont {Bertran}},
  \bibinfo {author} {\bibfnamefont {P.}~\bibnamefont {Le~F{\`e}vre}}, \bibinfo
  {author} {\bibfnamefont {A.}~\bibnamefont {Taleb-Ibrahimi}}, \bibinfo
  {author} {\bibfnamefont {M.-C.}\ \bibnamefont {Asensio}}, \bibinfo {author}
  {\bibfnamefont {M.}~\bibnamefont {Shi}}, \bibinfo {author} {\bibfnamefont
  {H.}~\bibnamefont {Lin}}, \bibinfo {author} {\bibfnamefont {M.}~\bibnamefont
  {Hoesch}}, \bibinfo {author} {\bibfnamefont {S.}~\bibnamefont {Oh}}, \ and\
  \bibinfo {author} {\bibfnamefont {M.~Z.}\ \bibnamefont {Hasan}},\ }\href
  {\doibase 10.1126/sciadv.1501692} {\bibfield  {journal} {\bibinfo  {journal}
  {Science Advances}\ }\textbf {\bibinfo {volume} {3}} (\bibinfo {year}
  {2017}),\ 10.1126/sciadv.1501692}\BibitemShut {NoStop}%
\bibitem [{\citenamefont {Agapito}\ \emph {et~al.}(2013)\citenamefont
  {Agapito}, \citenamefont {Ferretti}, \citenamefont {Calzolari}, \citenamefont
  {Curtarolo},\ and\ \citenamefont {Buongiorno~Nardelli}}]{PAO1}%
  \BibitemOpen
  \bibfield  {author} {\bibinfo {author} {\bibfnamefont {L.~A.}\ \bibnamefont
  {Agapito}}, \bibinfo {author} {\bibfnamefont {A.}~\bibnamefont {Ferretti}},
  \bibinfo {author} {\bibfnamefont {A.}~\bibnamefont {Calzolari}}, \bibinfo
  {author} {\bibfnamefont {S.}~\bibnamefont {Curtarolo}}, \ and\ \bibinfo
  {author} {\bibfnamefont {M.}~\bibnamefont {Buongiorno~Nardelli}},\ }\href
  {\doibase 10.1103/PhysRevB.88.165127} {\bibfield  {journal} {\bibinfo
  {journal} {Phys. Rev. B}\ }\textbf {\bibinfo {volume} {88}},\ \bibinfo
  {pages} {165127} (\bibinfo {year} {2013})}\BibitemShut {NoStop}%
\bibitem [{\citenamefont {Agapito}\ \emph {et~al.}(2015)\citenamefont
  {Agapito}, \citenamefont {Curtarolo},\ and\ \citenamefont
  {Buongiorno~Nardelli}}]{PAO2}%
  \BibitemOpen
  \bibfield  {author} {\bibinfo {author} {\bibfnamefont {L.~A.}\ \bibnamefont
  {Agapito}}, \bibinfo {author} {\bibfnamefont {S.}~\bibnamefont {Curtarolo}},
  \ and\ \bibinfo {author} {\bibfnamefont {M.}~\bibnamefont
  {Buongiorno~Nardelli}},\ }\href {\doibase 10.1103/PhysRevX.5.011006}
  {\bibfield  {journal} {\bibinfo  {journal} {Phys. Rev. X}\ }\textbf {\bibinfo
  {volume} {5}},\ \bibinfo {pages} {011006} (\bibinfo {year}
  {2015})}\BibitemShut {NoStop}%
\bibitem [{\citenamefont {Agapito}\ \emph
  {et~al.}(2016{\natexlab{a}})\citenamefont {Agapito}, \citenamefont {Fornari},
  \citenamefont {Ceresoli}, \citenamefont {Ferretti}, \citenamefont
  {Curtarolo},\ and\ \citenamefont {Nardelli}}]{PAO3}%
  \BibitemOpen
  \bibfield  {author} {\bibinfo {author} {\bibfnamefont {L.~A.}\ \bibnamefont
  {Agapito}}, \bibinfo {author} {\bibfnamefont {M.}~\bibnamefont {Fornari}},
  \bibinfo {author} {\bibfnamefont {D.}~\bibnamefont {Ceresoli}}, \bibinfo
  {author} {\bibfnamefont {A.}~\bibnamefont {Ferretti}}, \bibinfo {author}
  {\bibfnamefont {S.}~\bibnamefont {Curtarolo}}, \ and\ \bibinfo {author}
  {\bibfnamefont {M.~B.}\ \bibnamefont {Nardelli}},\ }\href {\doibase
  10.1103/PhysRevB.93.125137} {\bibfield  {journal} {\bibinfo  {journal} {Phys.
  Rev. B}\ }\textbf {\bibinfo {volume} {93}},\ \bibinfo {pages} {125137}
  (\bibinfo {year} {2016}{\natexlab{a}})}\BibitemShut {NoStop}%
\bibitem [{\citenamefont {Agapito}\ \emph
  {et~al.}(2016{\natexlab{b}})\citenamefont {Agapito}, \citenamefont
  {Ismail-Beigi}, \citenamefont {Curtarolo}, \citenamefont {Fornari},\ and\
  \citenamefont {Nardelli}}]{PAO4}%
  \BibitemOpen
  \bibfield  {author} {\bibinfo {author} {\bibfnamefont {L.~A.}\ \bibnamefont
  {Agapito}}, \bibinfo {author} {\bibfnamefont {S.}~\bibnamefont
  {Ismail-Beigi}}, \bibinfo {author} {\bibfnamefont {S.}~\bibnamefont
  {Curtarolo}}, \bibinfo {author} {\bibfnamefont {M.}~\bibnamefont {Fornari}},
  \ and\ \bibinfo {author} {\bibfnamefont {M.~B.}\ \bibnamefont {Nardelli}},\
  }\href {\doibase 10.1103/PhysRevB.93.035104} {\bibfield  {journal} {\bibinfo
  {journal} {Phys. Rev. B}\ }\textbf {\bibinfo {volume} {93}},\ \bibinfo
  {pages} {035104} (\bibinfo {year} {2016}{\natexlab{b}})}\BibitemShut
  {NoStop}%
\bibitem [{\citenamefont {Nardelli}\ \emph {et~al.}(2018)\citenamefont
  {Nardelli}, \citenamefont {Cerasoli}, \citenamefont {Costa}, \citenamefont
  {Curtarolo}, \citenamefont {Gennaro}, \citenamefont {Fornari}, \citenamefont
  {Liyanage}, \citenamefont {Supka},\ and\ \citenamefont {Wang}}]{PAO5}%
  \BibitemOpen
  \bibfield  {author} {\bibinfo {author} {\bibfnamefont {M.~B.}\ \bibnamefont
  {Nardelli}}, \bibinfo {author} {\bibfnamefont {F.~T.}\ \bibnamefont
  {Cerasoli}}, \bibinfo {author} {\bibfnamefont {M.}~\bibnamefont {Costa}},
  \bibinfo {author} {\bibfnamefont {S.}~\bibnamefont {Curtarolo}}, \bibinfo
  {author} {\bibfnamefont {R.~D.}\ \bibnamefont {Gennaro}}, \bibinfo {author}
  {\bibfnamefont {M.}~\bibnamefont {Fornari}}, \bibinfo {author} {\bibfnamefont
  {L.}~\bibnamefont {Liyanage}}, \bibinfo {author} {\bibfnamefont {A.~R.}\
  \bibnamefont {Supka}}, \ and\ \bibinfo {author} {\bibfnamefont
  {H.}~\bibnamefont {Wang}},\ }\href {\doibase
  https://doi.org/10.1016/j.commatsci.2017.11.034} {\bibfield  {journal}
  {\bibinfo  {journal} {Computational Materials Science}\ }\textbf {\bibinfo
  {volume} {143}},\ \bibinfo {pages} {462 } (\bibinfo {year}
  {2018})}\BibitemShut {NoStop}%
\bibitem [{\citenamefont {Corso}\ and\ \citenamefont {Conte}(2005)}]{DFT-SOC}%
  \BibitemOpen
  \bibfield  {author} {\bibinfo {author} {\bibfnamefont {A.~D.}\ \bibnamefont
  {Corso}}\ and\ \bibinfo {author} {\bibfnamefont {A.~M.}\ \bibnamefont
  {Conte}},\ }\href {\doibase 10.1103/PhysRevB.71.115106} {\bibfield  {journal}
  {\bibinfo  {journal} {Phys. Rev. B}\ }\textbf {\bibinfo {volume} {71}},\
  \bibinfo {pages} {115106} (\bibinfo {year} {2005})}\BibitemShut {NoStop}%
\bibitem [{\citenamefont {Abate}\ and\ \citenamefont {Asdente}(1965)}]{soc}%
  \BibitemOpen
  \bibfield  {author} {\bibinfo {author} {\bibfnamefont {E.}~\bibnamefont
  {Abate}}\ and\ \bibinfo {author} {\bibfnamefont {M.}~\bibnamefont
  {Asdente}},\ }\href {\doibase 10.1103/PhysRev.140.A1303} {\bibfield
  {journal} {\bibinfo  {journal} {Phys. Rev.}\ }\textbf {\bibinfo {volume}
  {140}},\ \bibinfo {pages} {A1303} (\bibinfo {year} {1965})}\BibitemShut
  {NoStop}%
\bibitem [{\citenamefont {Zhang}\ \emph {et~al.}(2009)\citenamefont {Zhang},
  \citenamefont {Liu}, \citenamefont {Qi}, \citenamefont {Dai}, \citenamefont
  {Fang},\ and\ \citenamefont {Zhang}}]{bi2se3}%
  \BibitemOpen
  \bibfield  {author} {\bibinfo {author} {\bibfnamefont {H.}~\bibnamefont
  {Zhang}}, \bibinfo {author} {\bibfnamefont {C.-X.}\ \bibnamefont {Liu}},
  \bibinfo {author} {\bibfnamefont {X.-L.}\ \bibnamefont {Qi}}, \bibinfo
  {author} {\bibfnamefont {X.}~\bibnamefont {Dai}}, \bibinfo {author}
  {\bibfnamefont {Z.}~\bibnamefont {Fang}}, \ and\ \bibinfo {author}
  {\bibfnamefont {S.-C.}\ \bibnamefont {Zhang}},\ }\href {\doibase
  10.1038/nphys1270} {\bibfield  {journal} {\bibinfo  {journal} {Nat Phys}\
  }\textbf {\bibinfo {volume} {5}},\ \bibinfo {pages} {438} (\bibinfo {year}
  {2009})}\BibitemShut {NoStop}%
\bibitem [{\citenamefont {Hohenberg}\ and\ \citenamefont {Kohn}(1964)}]{DFT1}%
  \BibitemOpen
  \bibfield  {author} {\bibinfo {author} {\bibfnamefont {P.}~\bibnamefont
  {Hohenberg}}\ and\ \bibinfo {author} {\bibfnamefont {W.}~\bibnamefont
  {Kohn}},\ }\href {\doibase 10.1103/PhysRev.136.B864} {\bibfield  {journal}
  {\bibinfo  {journal} {Phys. Rev.}\ }\textbf {\bibinfo {volume} {136}},\
  \bibinfo {pages} {B864} (\bibinfo {year} {1964})}\BibitemShut {NoStop}%
\bibitem [{\citenamefont {Kohn}\ and\ \citenamefont {Sham}(1965)}]{DFT2}%
  \BibitemOpen
  \bibfield  {author} {\bibinfo {author} {\bibfnamefont {W.}~\bibnamefont
  {Kohn}}\ and\ \bibinfo {author} {\bibfnamefont {L.~J.}\ \bibnamefont
  {Sham}},\ }\href {\doibase 10.1103/PhysRev.140.A1133} {\bibfield  {journal}
  {\bibinfo  {journal} {Phys. Rev.}\ }\textbf {\bibinfo {volume} {140}},\
  \bibinfo {pages} {A1133} (\bibinfo {year} {1965})}\BibitemShut {NoStop}%
\bibitem [{\citenamefont {Giannozzi}\ \emph {et~al.}(2009)\citenamefont
  {Giannozzi}, \citenamefont {Baroni}, \citenamefont {Bonini}, \citenamefont
  {Calandra}, \citenamefont {Car}, \citenamefont {Cavazzoni}, \citenamefont
  {Ceresoli}, \citenamefont {Chiarotti}, \citenamefont {Cococcioni},
  \citenamefont {Dabo}, \citenamefont {{Dal Corso}}, \citenamefont
  {de~Gironcoli}, \citenamefont {Fabris}, \citenamefont {Fratesi},
  \citenamefont {Gebauer}, \citenamefont {Gerstmann}, \citenamefont
  {Gougoussis}, \citenamefont {Kokalj}, \citenamefont {Lazzeri}, \citenamefont
  {Martin-Samos}, \citenamefont {Marzari}, \citenamefont {Mauri}, \citenamefont
  {Mazzarello}, \citenamefont {Paolini}, \citenamefont {Pasquarello},
  \citenamefont {Paulatto}, \citenamefont {Sbraccia}, \citenamefont {Scandolo},
  \citenamefont {Sclauzero}, \citenamefont {Seitsonen}, \citenamefont
  {Smogunov}, \citenamefont {Umari},\ and\ \citenamefont
  {Wentzcovitch}}]{QE-2009}%
  \BibitemOpen
  \bibfield  {author} {\bibinfo {author} {\bibfnamefont {P.}~\bibnamefont
  {Giannozzi}}, \bibinfo {author} {\bibfnamefont {S.}~\bibnamefont {Baroni}},
  \bibinfo {author} {\bibfnamefont {N.}~\bibnamefont {Bonini}}, \bibinfo
  {author} {\bibfnamefont {M.}~\bibnamefont {Calandra}}, \bibinfo {author}
  {\bibfnamefont {R.}~\bibnamefont {Car}}, \bibinfo {author} {\bibfnamefont
  {C.}~\bibnamefont {Cavazzoni}}, \bibinfo {author} {\bibfnamefont
  {D.}~\bibnamefont {Ceresoli}}, \bibinfo {author} {\bibfnamefont {G.~L.}\
  \bibnamefont {Chiarotti}}, \bibinfo {author} {\bibfnamefont {M.}~\bibnamefont
  {Cococcioni}}, \bibinfo {author} {\bibfnamefont {I.}~\bibnamefont {Dabo}},
  \bibinfo {author} {\bibfnamefont {A.}~\bibnamefont {{Dal Corso}}}, \bibinfo
  {author} {\bibfnamefont {S.}~\bibnamefont {de~Gironcoli}}, \bibinfo {author}
  {\bibfnamefont {S.}~\bibnamefont {Fabris}}, \bibinfo {author} {\bibfnamefont
  {G.}~\bibnamefont {Fratesi}}, \bibinfo {author} {\bibfnamefont
  {R.}~\bibnamefont {Gebauer}}, \bibinfo {author} {\bibfnamefont
  {U.}~\bibnamefont {Gerstmann}}, \bibinfo {author} {\bibfnamefont
  {C.}~\bibnamefont {Gougoussis}}, \bibinfo {author} {\bibfnamefont
  {A.}~\bibnamefont {Kokalj}}, \bibinfo {author} {\bibfnamefont
  {M.}~\bibnamefont {Lazzeri}}, \bibinfo {author} {\bibfnamefont
  {L.}~\bibnamefont {Martin-Samos}}, \bibinfo {author} {\bibfnamefont
  {N.}~\bibnamefont {Marzari}}, \bibinfo {author} {\bibfnamefont
  {F.}~\bibnamefont {Mauri}}, \bibinfo {author} {\bibfnamefont
  {R.}~\bibnamefont {Mazzarello}}, \bibinfo {author} {\bibfnamefont
  {S.}~\bibnamefont {Paolini}}, \bibinfo {author} {\bibfnamefont
  {A.}~\bibnamefont {Pasquarello}}, \bibinfo {author} {\bibfnamefont
  {L.}~\bibnamefont {Paulatto}}, \bibinfo {author} {\bibfnamefont
  {C.}~\bibnamefont {Sbraccia}}, \bibinfo {author} {\bibfnamefont
  {S.}~\bibnamefont {Scandolo}}, \bibinfo {author} {\bibfnamefont
  {G.}~\bibnamefont {Sclauzero}}, \bibinfo {author} {\bibfnamefont {A.~P.}\
  \bibnamefont {Seitsonen}}, \bibinfo {author} {\bibfnamefont {A.}~\bibnamefont
  {Smogunov}}, \bibinfo {author} {\bibfnamefont {P.}~\bibnamefont {Umari}}, \
  and\ \bibinfo {author} {\bibfnamefont {R.~M.}\ \bibnamefont {Wentzcovitch}},\
  }\href {http://www.quantum-espresso.org} {\bibfield  {journal} {\bibinfo
  {journal} {Journal of Physics: Condensed Matter}\ }\textbf {\bibinfo {volume}
  {21}},\ \bibinfo {pages} {395502 (19pp)} (\bibinfo {year}
  {2009})}\BibitemShut {NoStop}%
\bibitem [{\citenamefont {Giannozzi}\ \emph {et~al.}(2017)\citenamefont
  {Giannozzi}, \citenamefont {Andreussi}, \citenamefont {Brumme}, \citenamefont
  {Bunau}, \citenamefont {Nardelli}, \citenamefont {Calandra}, \citenamefont
  {Car}, \citenamefont {Cavazzoni}, \citenamefont {Ceresoli}, \citenamefont
  {Cococcioni}, \citenamefont {Colonna}, \citenamefont {Carnimeo},
  \citenamefont {Corso}, \citenamefont {de~Gironcoli}, \citenamefont {Delugas},
  \citenamefont {Jr}, \citenamefont {Ferretti}, \citenamefont {Floris},
  \citenamefont {Fratesi}, \citenamefont {Fugallo}, \citenamefont {Gebauer},
  \citenamefont {Gerstmann}, \citenamefont {Giustino}, \citenamefont {Gorni},
  \citenamefont {Jia}, \citenamefont {Kawamura}, \citenamefont {Ko},
  \citenamefont {Kokalj}, \citenamefont {Küçükbenli}, \citenamefont
  {Lazzeri}, \citenamefont {Marsili}, \citenamefont {Marzari}, \citenamefont
  {Mauri}, \citenamefont {Nguyen}, \citenamefont {Nguyen}, \citenamefont {de-la
  Roza}, \citenamefont {Paulatto}, \citenamefont {Poncé}, \citenamefont
  {Rocca}, \citenamefont {Sabatini}, \citenamefont {Santra}, \citenamefont
  {Schlipf}, \citenamefont {Seitsonen}, \citenamefont {Smogunov}, \citenamefont
  {Timrov}, \citenamefont {Thonhauser}, \citenamefont {Umari}, \citenamefont
  {Vast}, \citenamefont {Wu},\ and\ \citenamefont {Baroni}}]{QE-2017}%
  \BibitemOpen
  \bibfield  {author} {\bibinfo {author} {\bibfnamefont {P.}~\bibnamefont
  {Giannozzi}}, \bibinfo {author} {\bibfnamefont {O.}~\bibnamefont
  {Andreussi}}, \bibinfo {author} {\bibfnamefont {T.}~\bibnamefont {Brumme}},
  \bibinfo {author} {\bibfnamefont {O.}~\bibnamefont {Bunau}}, \bibinfo
  {author} {\bibfnamefont {M.~B.}\ \bibnamefont {Nardelli}}, \bibinfo {author}
  {\bibfnamefont {M.}~\bibnamefont {Calandra}}, \bibinfo {author}
  {\bibfnamefont {R.}~\bibnamefont {Car}}, \bibinfo {author} {\bibfnamefont
  {C.}~\bibnamefont {Cavazzoni}}, \bibinfo {author} {\bibfnamefont
  {D.}~\bibnamefont {Ceresoli}}, \bibinfo {author} {\bibfnamefont
  {M.}~\bibnamefont {Cococcioni}}, \bibinfo {author} {\bibfnamefont
  {N.}~\bibnamefont {Colonna}}, \bibinfo {author} {\bibfnamefont
  {I.}~\bibnamefont {Carnimeo}}, \bibinfo {author} {\bibfnamefont {A.~D.}\
  \bibnamefont {Corso}}, \bibinfo {author} {\bibfnamefont {S.}~\bibnamefont
  {de~Gironcoli}}, \bibinfo {author} {\bibfnamefont {P.}~\bibnamefont
  {Delugas}}, \bibinfo {author} {\bibfnamefont {R.~A.~D.}\ \bibnamefont {Jr}},
  \bibinfo {author} {\bibfnamefont {A.}~\bibnamefont {Ferretti}}, \bibinfo
  {author} {\bibfnamefont {A.}~\bibnamefont {Floris}}, \bibinfo {author}
  {\bibfnamefont {G.}~\bibnamefont {Fratesi}}, \bibinfo {author} {\bibfnamefont
  {G.}~\bibnamefont {Fugallo}}, \bibinfo {author} {\bibfnamefont
  {R.}~\bibnamefont {Gebauer}}, \bibinfo {author} {\bibfnamefont
  {U.}~\bibnamefont {Gerstmann}}, \bibinfo {author} {\bibfnamefont
  {F.}~\bibnamefont {Giustino}}, \bibinfo {author} {\bibfnamefont
  {T.}~\bibnamefont {Gorni}}, \bibinfo {author} {\bibfnamefont
  {J.}~\bibnamefont {Jia}}, \bibinfo {author} {\bibfnamefont {M.}~\bibnamefont
  {Kawamura}}, \bibinfo {author} {\bibfnamefont {H.-Y.}\ \bibnamefont {Ko}},
  \bibinfo {author} {\bibfnamefont {A.}~\bibnamefont {Kokalj}}, \bibinfo
  {author} {\bibfnamefont {E.}~\bibnamefont {Küçükbenli}}, \bibinfo {author}
  {\bibfnamefont {M.}~\bibnamefont {Lazzeri}}, \bibinfo {author} {\bibfnamefont
  {M.}~\bibnamefont {Marsili}}, \bibinfo {author} {\bibfnamefont
  {N.}~\bibnamefont {Marzari}}, \bibinfo {author} {\bibfnamefont
  {F.}~\bibnamefont {Mauri}}, \bibinfo {author} {\bibfnamefont {N.~L.}\
  \bibnamefont {Nguyen}}, \bibinfo {author} {\bibfnamefont {H.-V.}\
  \bibnamefont {Nguyen}}, \bibinfo {author} {\bibfnamefont {A.~O.}\
  \bibnamefont {de-la Roza}}, \bibinfo {author} {\bibfnamefont
  {L.}~\bibnamefont {Paulatto}}, \bibinfo {author} {\bibfnamefont
  {S.}~\bibnamefont {Poncé}}, \bibinfo {author} {\bibfnamefont
  {D.}~\bibnamefont {Rocca}}, \bibinfo {author} {\bibfnamefont
  {R.}~\bibnamefont {Sabatini}}, \bibinfo {author} {\bibfnamefont
  {B.}~\bibnamefont {Santra}}, \bibinfo {author} {\bibfnamefont
  {M.}~\bibnamefont {Schlipf}}, \bibinfo {author} {\bibfnamefont {A.~P.}\
  \bibnamefont {Seitsonen}}, \bibinfo {author} {\bibfnamefont {A.}~\bibnamefont
  {Smogunov}}, \bibinfo {author} {\bibfnamefont {I.}~\bibnamefont {Timrov}},
  \bibinfo {author} {\bibfnamefont {T.}~\bibnamefont {Thonhauser}}, \bibinfo
  {author} {\bibfnamefont {P.}~\bibnamefont {Umari}}, \bibinfo {author}
  {\bibfnamefont {N.}~\bibnamefont {Vast}}, \bibinfo {author} {\bibfnamefont
  {X.}~\bibnamefont {Wu}}, \ and\ \bibinfo {author} {\bibfnamefont
  {S.}~\bibnamefont {Baroni}},\ }\href
  {http://stacks.iop.org/0953-8984/29/i=46/a=465901} {\bibfield  {journal}
  {\bibinfo  {journal} {Journal of Physics: Condensed Matter}\ }\textbf
  {\bibinfo {volume} {29}},\ \bibinfo {pages} {465901} (\bibinfo {year}
  {2017})}\BibitemShut {NoStop}%
\bibitem [{\citenamefont {Kresse}\ and\ \citenamefont
  {Furthm\"uller}(1996)}]{vasp1}%
  \BibitemOpen
  \bibfield  {author} {\bibinfo {author} {\bibfnamefont {G.}~\bibnamefont
  {Kresse}}\ and\ \bibinfo {author} {\bibfnamefont {J.}~\bibnamefont
  {Furthm\"uller}},\ }\href {\doibase
  https://doi.org/10.1016/0927-0256(96)00008-0} {\bibfield  {journal} {\bibinfo
   {journal} {Computational Materials Science}\ }\textbf {\bibinfo {volume}
  {6}},\ \bibinfo {pages} {15 } (\bibinfo {year} {1996})}\BibitemShut {NoStop}%
\bibitem [{\citenamefont {Kresse}\ and\ \citenamefont
  {Furthmuller}(1996)}]{vasp2}%
  \BibitemOpen
  \bibfield  {author} {\bibinfo {author} {\bibfnamefont {G.}~\bibnamefont
  {Kresse}}\ and\ \bibinfo {author} {\bibfnamefont {J.}~\bibnamefont
  {Furthmuller}},\ }\href {\doibase 10.1103/PhysRevB.54.11169} {\bibfield
  {journal} {\bibinfo  {journal} {Phys. Rev. B}\ }\textbf {\bibinfo {volume}
  {54}},\ \bibinfo {pages} {11169} (\bibinfo {year} {1996})}\BibitemShut
  {NoStop}%
\bibitem [{\citenamefont {Perdew}\ \emph {et~al.}(1996)\citenamefont {Perdew},
  \citenamefont {Burke},\ and\ \citenamefont {Ernzerhof}}]{pbe}%
  \BibitemOpen
  \bibfield  {author} {\bibinfo {author} {\bibfnamefont {J.~P.}\ \bibnamefont
  {Perdew}}, \bibinfo {author} {\bibfnamefont {K.}~\bibnamefont {Burke}}, \
  and\ \bibinfo {author} {\bibfnamefont {M.}~\bibnamefont {Ernzerhof}},\ }\href
  {\doibase 10.1103/PhysRevLett.77.3865} {\bibfield  {journal} {\bibinfo
  {journal} {Phys. Rev. Lett.}\ }\textbf {\bibinfo {volume} {77}},\ \bibinfo
  {pages} {3865} (\bibinfo {year} {1996})}\BibitemShut {NoStop}%
\bibitem [{\citenamefont {Kresse}\ and\ \citenamefont {Joubert}(1999)}]{PAW}%
  \BibitemOpen
  \bibfield  {author} {\bibinfo {author} {\bibfnamefont {G.}~\bibnamefont
  {Kresse}}\ and\ \bibinfo {author} {\bibfnamefont {D.}~\bibnamefont
  {Joubert}},\ }\href {\doibase 10.1103/PhysRevB.59.1758} {\bibfield  {journal}
  {\bibinfo  {journal} {Phys. Rev. B}\ }\textbf {\bibinfo {volume} {59}},\
  \bibinfo {pages} {1758} (\bibinfo {year} {1999})}\BibitemShut {NoStop}%
\bibitem [{\citenamefont {Nakayama}\ \emph {et~al.}(2012)\citenamefont
  {Nakayama}, \citenamefont {Eto}, \citenamefont {Tanaka}, \citenamefont
  {Sato}, \citenamefont {Souma}, \citenamefont {Takahashi}, \citenamefont
  {Segawa},\ and\ \citenamefont {Ando}}]{NatHet}%
  \BibitemOpen
  \bibfield  {author} {\bibinfo {author} {\bibfnamefont {K.}~\bibnamefont
  {Nakayama}}, \bibinfo {author} {\bibfnamefont {K.}~\bibnamefont {Eto}},
  \bibinfo {author} {\bibfnamefont {Y.}~\bibnamefont {Tanaka}}, \bibinfo
  {author} {\bibfnamefont {T.}~\bibnamefont {Sato}}, \bibinfo {author}
  {\bibfnamefont {S.}~\bibnamefont {Souma}}, \bibinfo {author} {\bibfnamefont
  {T.}~\bibnamefont {Takahashi}}, \bibinfo {author} {\bibfnamefont
  {K.}~\bibnamefont {Segawa}}, \ and\ \bibinfo {author} {\bibfnamefont
  {Y.}~\bibnamefont {Ando}},\ }\href {\doibase 10.1103/PhysRevLett.109.236804}
  {\bibfield  {journal} {\bibinfo  {journal} {Phys. Rev. Lett.}\ }\textbf
  {\bibinfo {volume} {109}},\ \bibinfo {pages} {236804} (\bibinfo {year}
  {2012})}\BibitemShut {NoStop}%
\bibitem [{\citenamefont {Jin}\ \emph {et~al.}(2016)\citenamefont {Jin},
  \citenamefont {Yeom},\ and\ \citenamefont {Jhi}}]{band-alignment}%
  \BibitemOpen
  \bibfield  {author} {\bibinfo {author} {\bibfnamefont {K.-H.}\ \bibnamefont
  {Jin}}, \bibinfo {author} {\bibfnamefont {H.~W.}\ \bibnamefont {Yeom}}, \
  and\ \bibinfo {author} {\bibfnamefont {S.-H.}\ \bibnamefont {Jhi}},\ }\href
  {\doibase 10.1103/PhysRevB.93.075308} {\bibfield  {journal} {\bibinfo
  {journal} {Phys. Rev. B}\ }\textbf {\bibinfo {volume} {93}},\ \bibinfo
  {pages} {075308} (\bibinfo {year} {2016})}\BibitemShut {NoStop}%
\bibitem [{\citenamefont {Yazyev}\ \emph {et~al.}(2010)\citenamefont {Yazyev},
  \citenamefont {Moore},\ and\ \citenamefont {Louie}}]{bulk-states1}%
  \BibitemOpen
  \bibfield  {author} {\bibinfo {author} {\bibfnamefont {O.~V.}\ \bibnamefont
  {Yazyev}}, \bibinfo {author} {\bibfnamefont {J.~E.}\ \bibnamefont {Moore}}, \
  and\ \bibinfo {author} {\bibfnamefont {S.~G.}\ \bibnamefont {Louie}},\ }\href
  {\doibase 10.1103/PhysRevLett.105.266806} {\bibfield  {journal} {\bibinfo
  {journal} {Phys. Rev. Lett.}\ }\textbf {\bibinfo {volume} {105}},\ \bibinfo
  {pages} {266806} (\bibinfo {year} {2010})}\BibitemShut {NoStop}%
\bibitem [{\citenamefont {{Mera Acosta}}\ \emph {et~al.}(2018)\citenamefont
  {{Mera Acosta}}, \citenamefont {{Lima}}, \citenamefont {{da Silva}},
  \citenamefont {{Fazzio}},\ and\ \citenamefont {{Lewenkopf}}}]{bulk-states2}%
  \BibitemOpen
  \bibfield  {author} {\bibinfo {author} {\bibfnamefont {C.}~\bibnamefont
  {{Mera Acosta}}}, \bibinfo {author} {\bibfnamefont {M.~P.}\ \bibnamefont
  {{Lima}}}, \bibinfo {author} {\bibfnamefont {A.~J.~R.}\ \bibnamefont {{da
  Silva}}}, \bibinfo {author} {\bibfnamefont {A.}~\bibnamefont {{Fazzio}}}, \
  and\ \bibinfo {author} {\bibfnamefont {C.~H.}\ \bibnamefont {{Lewenkopf}}},\
  }\href@noop {} {\bibfield  {journal} {\bibinfo  {journal} {ArXiv e-prints}\ }
  (\bibinfo {year} {2018})},\ \Eprint {http://arxiv.org/abs/1802.07864}
  {arXiv:1802.07864 [cond-mat.mes-hall]} \BibitemShut {NoStop}%
\bibitem [{\citenamefont {Abdalla}\ \emph
  {et~al.}(2015{\natexlab{a}})\citenamefont {Abdalla}, \citenamefont {José},
  \citenamefont {Schmidt}, \citenamefont {Miwa},\ and\ \citenamefont
  {Fazzio}}]{abdalla}%
  \BibitemOpen
  \bibfield  {author} {\bibinfo {author} {\bibfnamefont {L.~B.}\ \bibnamefont
  {Abdalla}}, \bibinfo {author} {\bibfnamefont {E.~P.}\ \bibnamefont {José}},
  \bibinfo {author} {\bibfnamefont {T.~M.}\ \bibnamefont {Schmidt}}, \bibinfo
  {author} {\bibfnamefont {R.~H.}\ \bibnamefont {Miwa}}, \ and\ \bibinfo
  {author} {\bibfnamefont {A.}~\bibnamefont {Fazzio}},\ }\href
  {http://stacks.iop.org/0953-8984/27/i=25/a=255501} {\bibfield  {journal}
  {\bibinfo  {journal} {Journal of Physics: Condensed Matter}\ }\textbf
  {\bibinfo {volume} {27}},\ \bibinfo {pages} {255501} (\bibinfo {year}
  {2015}{\natexlab{a}})}\BibitemShut {NoStop}%
\bibitem [{\citenamefont {Shan}\ \emph {et~al.}(2010)\citenamefont {Shan},
  \citenamefont {Lu},\ and\ \citenamefont {Shen}}]{Shan-2010}%
  \BibitemOpen
  \bibfield  {author} {\bibinfo {author} {\bibfnamefont {W.-Y.}\ \bibnamefont
  {Shan}}, \bibinfo {author} {\bibfnamefont {H.-Z.}\ \bibnamefont {Lu}}, \ and\
  \bibinfo {author} {\bibfnamefont {S.-Q.}\ \bibnamefont {Shen}},\ }\href
  {http://stacks.iop.org/1367-2630/12/i=4/a=043048} {\bibfield  {journal}
  {\bibinfo  {journal} {New Journal of Physics}\ }\textbf {\bibinfo {volume}
  {12}},\ \bibinfo {pages} {043048} (\bibinfo {year} {2010})}\BibitemShut
  {NoStop}%
\bibitem [{\citenamefont {Abdalla}\ \emph
  {et~al.}(2015{\natexlab{b}})\citenamefont {Abdalla}, \citenamefont {Padilha},
  \citenamefont {Schmidt},\ and\ \citenamefont {Fazzio}}]{Abdalla-2015}%
  \BibitemOpen
  \bibfield  {author} {\bibinfo {author} {\bibfnamefont {L.~B.}\ \bibnamefont
  {Abdalla}}, \bibinfo {author} {\bibfnamefont {J.~E.}\ \bibnamefont
  {Padilha}}, \bibinfo {author} {\bibfnamefont {T.~M.}\ \bibnamefont
  {Schmidt}}, \ and\ \bibinfo {author} {\bibfnamefont {A.}~\bibnamefont
  {Fazzio}},\ }\href {\doibase 10.1088/0953-8984/27/25/255501} {\bibfield
  {journal} {\bibinfo  {journal} {Journal of Physics: Condensed Matter}\
  }\textbf {\bibinfo {volume} {27}},\ \bibinfo {pages} {255501} (\bibinfo
  {year} {2015}{\natexlab{b}})}\BibitemShut {NoStop}%
\end{thebibliography}%

\end{document}